\documentclass[a4paper,11pt]{article}
\usepackage{aaskaiid}
\usepackage{xcolor}
\usepackage{enumitem}
\usepackage{hyperref,upgreek}
\usepackage{multirow}
\usepackage{orcidlink}
\usepackage[utf8]{inputenc}

\title{High-Redshift Signatures from the Cosmic Dawn and the Epoch of Reionization}
\ShortTitle{High-Redshift Signatures}

\ShortName{Barkana et al.} 
\author[1]{Rennan Barkana\orcidlink{0000-0002-1557-693X}}
\author[17]{Oliver Basquette\orcidlink{0009-0006-4922-0209}}
\author[2]{Ankita Bera\orcidlink{0000-0001-7072-570X}}
\author[19, 20, 21]{Jennifer Yik Ham Chan\orcidlink{0000-0003-0314-7027}}
\author[24]{Pravabati Chingangbam\orcidlink{0000-0002-7385-8273}}
\author[3]{Hector Afonso G. Cruz\orcidlink{0000-0002-1775-3602}}
\author[17]{Saswata Dasgupta\orcidlink{0000-0001-6461-769X}}
\author[4]{Kanan K. Datta\orcidlink{0000-0002-2238-5146}}
\author[17, 18]{Anastasia Fialkov\orcidlink{0000-0002-1369-633X}}
\author[4,5]{Sambit K. Giri\orcidlink{0000-0002-2560-536X}}
\author[22]{Qin Han}
\author[6]{Ilian T. Iliev}
\author[7]{Bohua Li\orcidlink{0000-0002-3600-0358}}
\author[8]{Teppei Minoda\orcidlink{0000-0001-8333-2809}}
\author[9,10]{Shikhar Mittal\orcidlink{0000-0002-0247-618X}}
\author[11]{Julian B. Mu\~noz}
\author[23, 24]{Suvedha Suresh Naik\orcidlink{0000-0001-5272-966X}}
\author[16]{Janakee Raste\orcidlink{0000-0001-7451-6139}}
\author[12]{Aurel Schneider\orcidlink{0000-0001-7055-8104}}
\author[1]{Sudipta Sikder\orcidlink{0000-0001-6129-0118}}
\author[22]{Kinwah Wu\orcidlink{0000-0002-7568-8765}}
\author[13,14]{Yidong Xu\orcidlink{0000-0003-3224-4125}}
\author[13,14]{Bin Yue}
\author[13]{Meng Zhang}
\author[15]{Meng-Lin Zhao}

\affiliation[1]{School of Physics and Astronomy, Tel Aviv University, Tel Aviv, 69978, Israel}
\affiliation[2]{Department of Astronomy, University of Maryland, College Park, MD 20742, USA}
\affiliation[3]{Center for Cosmology and Particle Physics, Department of Physics, New York University, New York, NY 10003, USA}
\affiliation[4]{Department of Physics, Jadavpur University, 188, Raja S.C. Mallick Rd, Kolkata 700032, India}
\affiliation[4]{Department of Astronomy and Oskar Klein Centre, AlbaNova, Stockholm University, SE-10691 Stockholm, Sweden}
\affiliation[5]{Van Swinderen Institute for Particle Physics and Gravity, University of Groningen, Nijenborgh 3, 9747 AG Groningen, The Netherlands}
\affiliation[6]{Astronomy Centre, Department of Physics \& Astronomy, Pevensey III Building, University of Sussex, Falmer, Brighton, BN1 9QH, United Kingdom}
\affiliation[7]{Guangxi Key Laboratory for Relativistic Astrophysics, 
School of Physical Science and Technology, Guangxi University,
Nanning 530004, People’s Republic of China}
\affiliation[8]{Department of Astronomy, Tsinghua University, Beijing 100084, 
People’s Republic of China}
\affiliation[9]{Battcock Centre for Experimental Astrophysics, Cavendish Laboratory, J.~J.\ Thomson Avenue, Cambridge CB3 0HE, UK}
\affiliation[10]{Kavli Institute for Cosmology, University of Cambridge, Madingley Road, Cambridge CB3 0HA, UK}
\affiliation[11]{Department of Astronomy, The University of Texas at Austin, 2515 Speedway, Stop C1400, Austin, Texas 78712, USA}
\affiliation[12]{Department of Astrophysics, University of Zurich, Winterthurerstrasse 190, 8057 Zurich, Switzerland}
\affiliation[13]{National Astronomical Observatories, Chinese Academy of Sciences, Beijing 100101, People's Republic of China}
\affiliation[14]{State Key Laboratory of Radio Astronomy and Technology, Beijing 100101, People's Republic of China}
\affiliation[15]{Liaoning Key Laboratory of Cosmology and Astrophysics,
College of Sciences, Northeastern University, Shenyang 110819, People's Republic of China}
\affiliation[16]{National Centre for Radio Astrophysics, Tata Institute of Fundamental Research, Pune 411007, India}
\affiliation[17]{Institute of Astronomy, University of Cambridge, Madingley Road, Cambridge, CB3 0HA, UK}
\affiliation[18]{Kavli Institute for Cosmology, Madingley Road, Cambridge, CB3 0HA, UK}

\affiliation[19]{Department of Physics and Astronomy, Oberlin College, Oberlin, OH 44074, USA}
\affiliation[20]{Canadian Institute for Theoretical Astrophysics, University of Toronto, 60 St George St, Toronto, ON M5S 3H8, Canada}
\affiliation[21]{Dunlap Institute for Astronomy and Astrophysics, University of Toronto, 50 St George St, Toronto, ON M5S 3H8, Canada} 
\affiliation[22]{Mullard Space Science Laboratory, University College London, Holmbury St Mary, Surrey, RH5 6NT, UK}
\affiliation[23]{Korea Institute for Advanced Study (KIAS), 85 Hoegiro, Dongdaemun-gu, Seoul, Republic of Korea-02455}
\affiliation[24]{Indian Institute of Astrophysics, Koramangala II Block, Bangalore 560 034, India}

\abstract{In this chapter, we provide a comprehensive overview of the astrophysical and cosmological processes that shape the 21-cm signal during Cosmic Dawn and the Epoch of Reionization. We investigate both standard and exotic signatures potentially observable with SKA-Low. Standard signatures are those expected within the $\Lambda$CDM framework, including contributions from the first stars, galaxies, and black holes. Exotic signatures are more speculative indicating new physics, such as primordial black holes, modifications to the dark matter sector, non-standard primordial fluctuations, or strongly emitting radio galaxies. The effects of these different sources or scenarios are evaluated in the context of the expected sensitivity of SKA-Low, considering the AA* and AA4 configurations. The chapter aims to provide an overview of the theoretical landscape of 21-cm signatures and to highlight how the forthcoming SKA-Low observations will improve our understanding of astrophysical processes at early times and may open the door towards new physics beyond the $\Lambda$CDM framework.}

\begin{document}
\maketitle


\section{Introduction}
The redshifted 21-cm signal from neutral hydrogen has emerged as one of the most promising probes of the high-redshift Universe, offering a unique observational window into the epoch of reionization and the preceding cosmic dawn. During this period, the first generations of stars and galaxies emerged, leaving imprints on the intergalactic medium (IGM) that encode both astrophysical processes and potentially new physics.

The evolution of the signal is governed by three main drivers. First, the onset of Lyman-$\upalpha$ (Ly$\upalpha$) radiation from the earliest stars couples the hydrogen spin temperature to the gas kinetic temperature via the Wouthuysen-Field effect, making the IGM visible in absorption against the cosmic microwave background (CMB). Second, high-energy X-rays from the first stellar remnants and accreting black holes heat the IGM, transforming the absorption signal into emission as the gas temperature rises above the background. Finally, the growth of ionized regions around galaxies progressively erases the 21-cm signal fluctuations, culminating in the completion of reionization.

Together, these processes produce a rich  spectral and spatial signal that not only trace the birth and growth of structure in the Universe, but also provide an indirect laboratory for fundamental physics. It can be used to investigate extensions to the current cosmological model, including the nature of dark matter, exotic energy injection mechanisms, or new interactions beyond the Standard Model. As such, the 21-cm signal is a uniquely powerful probe, bridging astrophysics and fundamental physics at a time when the Universe was less than a billion years old.

The low frequency interferometer of the Square Kilometre Array Observatory (SKA-Low) will provide unprecedented measurements of the frequency range of 50-350 MHz, thereby observing the 21-cm signal at the redshifts $z\sim 6-30$. SKA-Low will primarily constrain the signal through power spectrum measurements, allowing the detection of brightness temperature fluctuations at the level of a few milli-Kelvin across a wide range of scales. Forecasts indicate that after a few thousand hours of integration, SKA-Low should be capable of measuring the 21-cm power spectrum with percent-level precision on scales of $0.1$-$1\,h$/Mpc during much of the reionization period.

Beyond statistical detections, the high angular resolution and sensitivity of SKA-Low will open the door to tomographic imaging of ionized regions in the mid and late stages of reionization. This would allow direct visualization of the ``bubble'' morphology driven by early galaxies and quasars, thereby connecting the astrophysical drivers of reionization to the evolving large-scale structure. In combination, these capabilities mean that SKA-Low will transform the study of the high-redshift Universe: delivering not only a detailed astrophysical narrative of cosmic dawn, but also placing tight constraints on exotic physics.


In this chapter we first summarise the physics of the 21-cm signal in the context of SKA-Low (Sec. \ref{sec:21cmsignal} and \ref{sec:SKALow}). We then go on and discuss both signatures from standard, astrophysical (Sec.~\ref{sec:StandardSignatures}) and from exotic origin (Sec.~\ref{sec:ExoticSignatures}) involving physics beyond the standard model. The standard signatures include early galaxies, Population III stars, X-ray binaries, Cosmic rays, and supermassive black holes. The section furthermore includes a discussion about velocity acoustic oscillations and the polarization of 21-cm radiation. In the exotic signature section we discuss strongly-emitting radio galaxies, primordial magnetic fields, baryon-dark matter interactions, dark matter annihilation and decay, non-cold dark matter, and primordial features from inflation. Throughout the whole chapter, we focus on the detectability of all these signatures assuming the AA* and AA4 configurations of the SKA-Low telescope.

\section{The physics of the 21-cm signal}
\label{sec:21cmsignal}
We start by providing a summary of the key processes shaping the 21-cm signal and a brief summary of how they are being modelled.

\begin{figure*}
    \centering
    \includegraphics[width=0.9\linewidth,trim=0.0cm 0.3cm 0.0cm 0.3cm, clip]{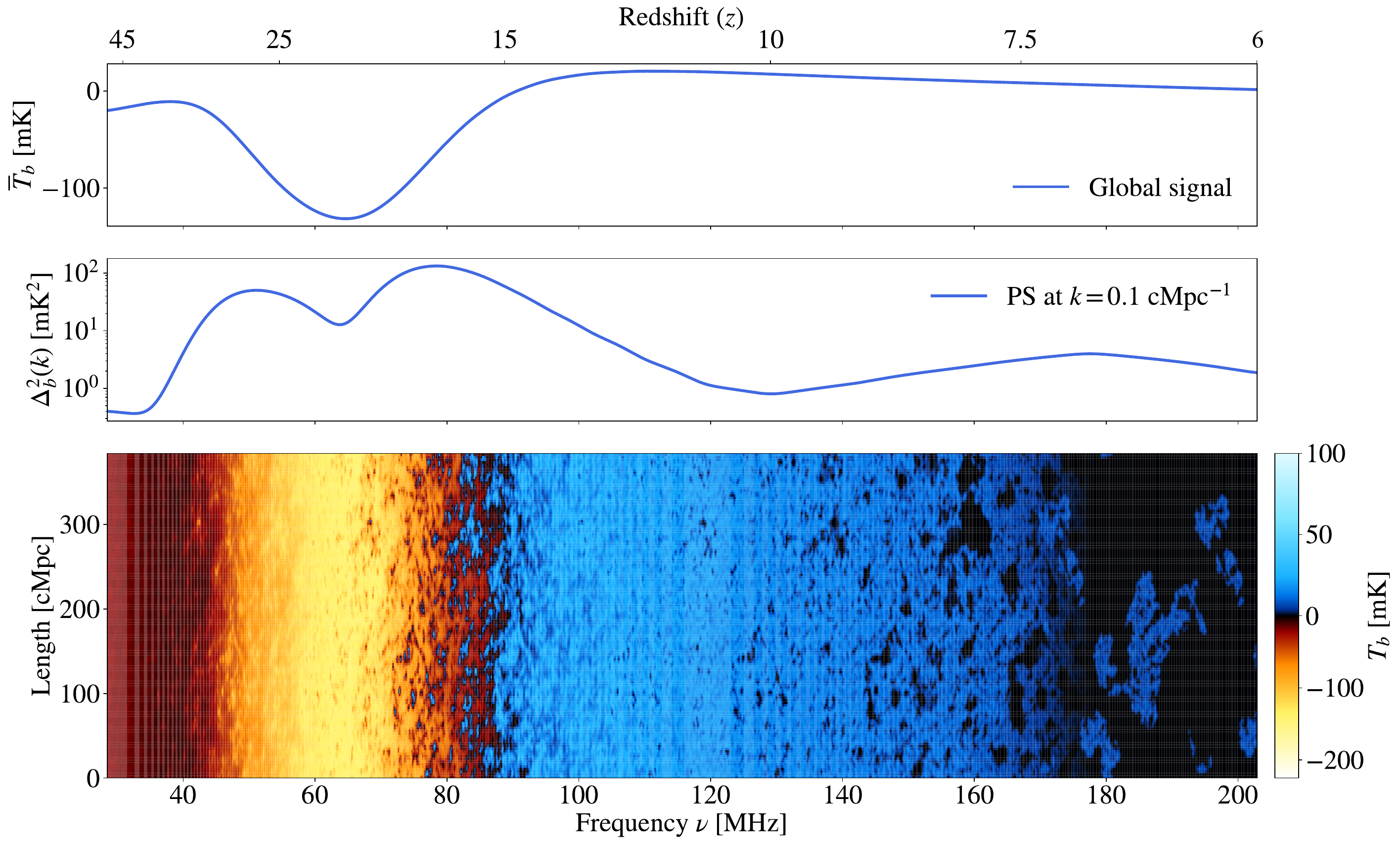}
        \caption{The 21-cm signal across cosmic epochs. The figure illustrates the evolution of the 21-cm signal from the Dark Ages through Cosmic Dawn to the end of the Epoch of Reionization, with reionization completed by $z \sim 6$. \textbf{Top}: The sky-averaged (global) 21-cm brightness temperature, $\overline{T}_{b}$, as a function of redshift (top x-axis)/frequency (bottom x-axis). The absorption trough at $z \sim 17$ reflects the onset of strong Ly$\upalpha$ coupling, followed by X-ray heating and reionization. \textbf{Middle}: The 21-cm power spectrum amplitude at $k = 0.1\ \mathrm{cMpc^{-1}}$ showing peaks associated with Ly$\upalpha$ coupling, X-ray heating, and ionization fluctuations. \textbf{Bottom}: A lightcone map of the differential brightness temperature, $T_{b}$, showing the spatial and temporal evolution of the signal. The signal is generated using the code \texttt{21cmSPACE} \citep[e.g.][]{Gessey2025}.
        }
        \label{fig:lightcone-global-ps}
\end{figure*}

\subsection{Lyman-\texorpdfstring{$\upalpha$}{α} coupling}
\label{sec:Lyacoupling}
During Cosmic Dawn, a strong 21-cm signal is expected in absorption against the CMB as a result of the Wouthuysen-Field (WF) effect \citep{Wouth, 1958PIRE...46..240F} caused by Ly$\upalpha$ photons from the first stars. The WF effect refers to a change in the occupation number of hyperfine states due to resonance scattering of Ly$\upalpha$ photons by the hydrogen atom. 
Stated differently, an electron excited by a Ly$\upalpha$ photon may return to a different hyperfine level it originally started from.

Before cosmic dawn, the collisional coupling was the dominant process leading to an absorption signal. As the universe expands, this becomes less efficient and, consequently, there is no contrast with the CMB, which results in a vanishing global signal, as seen at redshift $z\sim35$ in Fig.~\ref{fig:lightcone-global-ps} (top panel). Once the first sources turn on, their Ly$\upalpha$ radiation once again couples the 21-cm signal to the still-cold gas, resulting in an absorption trough as seen in Fig.~\ref{fig:lightcone-global-ps} at $z\sim25-15$. 

The first galaxies are rare and their spatial distribution initially exhibits strong Poisson fluctuations and is later highly clustered, resulting in strong fluctuations in the Ly$\upalpha$ background and thus the redshifted 21-cm line \citep{2005ApJ...626....1B}. This is evident from the middle panel in Fig.~\ref{fig:lightcone-global-ps} which shows the 21-cm power spectrum ($\Delta^2$) at a representative scale of $k=0.1\,\mathrm{cMpc^{-1}}$. 

The Ly$\upalpha$ coupling $x_{\mathrm{Ly}}$ depends on the local specific intensity $J_\upalpha$ which is expensive to compute. This is due to the large number of scatterings $\left(\sim\,10^6\right)$ occurring in the line core \citep{Loeb_1999, Baek_2009} at sub-Mpc scales which must be captured for an accurate model of $J_\upalpha$ and hence the Cosmic Dawn signal. Several different approaches have been proposed to deal with this, from analytical ones based on the wing-approximation \citep{Chuzhoy, FP06, Mesinger_11, Mittal_lya}, semi-numerical ones, where the radial profile ($J=J(\nu, r,z)$) for an idealised source assuming a spherical symmetry is pasted over all the sources in the cosmological box \citep{Reis_2021, Schaeffer_2023}, to numerical ones, where a hydrodynamical simulation is post-processed with Monte Carlo radiative transfer simulation \citep{Semelin23, Mittal_mcrt}.

It is now well-established that the lack of detailed Ly$\upalpha$ radiative transfer can result in modelling biases in the global 21-cm signal and even more so in calculating the 21-cm power spectrum \citep{Reis_2021, Semelin23, Mittal_mcrt}. Fluctuations in the Ly$\upalpha$ field for a wide range of scales can alter the 21-cm power spectrum by factor of up to a few. For example, consider an idealised Universe with no ionizing or X-ray sources so that the only sources of fluctuations are gas temperature and Ly$\upalpha$ coupling. At $k=0.07\,\mathrm{cMpc^{-1}}h$, we have $\Delta^2_{21}\approx 217\,\mathrm{mK^2}$ when a rigorous radiative transfer simulation of Ly$\upalpha$ is followed and $\Delta^2_{21}\approx 60\,\mathrm{mK^2}$ at the same wavenumber for no detailed radiative transfer \citep{Mittal_mcrt}. This stronger power spectrum is induced by the additional fluctuation in Ly$\upalpha$ coupling. In the middle panel of Fig.~\ref{fig:lightcone-global-ps}, the highest redshift peak in the 21-cm power spectrum is caused by fluctuations in $J_\upalpha$. Upcoming experiments such as \textit{SKAO} are strong candidates for verifying the 21-cm power spectrum for a wide range of scales.

\subsection{Heating process}
\label{sec:heating}
The thermal history of the IGM is a crucial component that drives the behaviour of the 21-cm signal on par with the Ly$\upalpha$ coupling (Sec.~\ref{sec:Lyacoupling}) and the process of ionization (discussed in Sec.~\ref{sec:ion_process}). Heating and cooling processes affect the 21-cm signal across the Dark Ages, Cosmic Dawn and the Epoch of Reionization. After recombination, the Universe cools largely adiabatically due to the cosmic expansion. Once it decouples thermally from the CMB at $z\sim 150$ \citep{Seager_1999}, the pristine gas, composed primarily of neutral hydrogen and helium, cools more rapidly than the CMB. This results in the IGM being colder than the CMB, which manifests as absorption by hydrogen atoms from the CMB, as shown in the top panel of Fig.~\ref{fig:lightcone-global-ps}. The subsequent thermal evolution, however, is shaped by a complex interplay of various other heating and cooling mechanisms that add to this expansion-driven cooling \citep[e.g.][]{MMR1997, F06, Pritchard_2007, Baek09, Mesinger_11, Fialkov_2014, Pacucci2014, Reis_2021, Gessey_2023}.  

Once the first collapsed structures begin to form, several heating mechanisms become important \citep{1992ApJ...386..432K,1997NewA....2..209A}. Shock heating arises from the gravitational collapse of matter, where supersonic gas flows create shocks that thermalize kinetic energy into heat \citep{Furlanetto_2004}. As the first stars form and die, they contribute several additional large-scale heating mechanisms mediated by X-rays \citep{Pritchard_2007}, cosmic rays \citep{Sazonov_2015, Bera_2023, Gessey_2023}, and Ly$\upalpha$ photons \citep{Chuzhoy2007,Mittal_lya, Reis_2021}. The Compton heating and cooling from scattering of CMB photons also contributes \citep{1992ApJ...386..432K}. The first prediction of observable 21-cm fluctuations from inhomogeneous X-ray heating was made by \citet{Pritchard_2007}, applying galaxy fluctuations to X-ray emission  \citep[instead of Ly$\upalpha$ emission which was first discussed by][]{2005ApJ...626....1B}.

At cosmic dawn, X-rays are thought to be produced by X-ray binaries \citep[XRBs][]{Power_2013, 2013ApJ...776L..31F} -- binary systems where matter accretes onto a compact object (neutron star or black hole) emitting copious X-rays (see further discussion in section \ref{sec:XRBs}),  and quasars \citep[likely subdominant at the redshifts relevant for the \textit{SKAO},][]{Madau_2004}. Observations suggest that the bolometric luminosity of X-rays, produced by a population of XRBs, scales linearly with the star formation rate \citep{2003MNRAS.339..793G, 2012MNRAS.419.2095M, 2014MNRAS.437.1698M} -- the dependence adopted in numerical simulations and semi-analytical modelling \citep{Pritchard_2007, 
Santos_2008, Mesinger_11, Fialkov_2014, 2017MNRAS.468.3785R, Eide2018}. In addition to the total intensity, the shape of the X-ray spectral energy distribution (SED) is an important property \citep[e.g.][]{Fialkov_2014,2017MNRAS.468.3785R}, as the mean free paths of X-rays in neutral IGM depend on their energy. X-rays with energies between $\sim 0.5-2$ keV travel great distances, contributing to the large-scale heating of the IGM (e.g. the mean free path of 1 keV photons in the neutral IGM at $z=14$ is $\approx180$ comoving Mpc). These X-rays also contribute to large-scale patterns of partial ionization in the IGM \citep[up to 10\%,][]{Baek10, Pacucci2014, Fialkov_2017}. Lower-energy photons ($\lesssim 0.5$ keV) are absorbed locally at the source, while the more energetic X-rays are not absorbed at all, thus contributing to the cosmic X-ray background \citep[e.g.][]{2013ApJ...776L..31F, Fialkov_2017, Ma2018}. 

Another proposed heating source, which competes with X-rays, is heating by cosmic rays, primarily produced by supernovae within the first galaxies \citep{Sazonov_2015, Bera_2023, Gessey_2023}. While cosmic rays have shorter mean free paths compared to X-rays and deposit their energy very close to the sources, cosmic rays can still contribute significantly to local heating, particularly in the vicinity of star-forming regions. For the further discussion of cosmic ray heating see section \ref{sec:CRs}.

One more heating mechanism that is crucial at cosmic dawn is the Ly$\upalpha$ heating. Along with mediating the coupling of spin temperature $T_{\rm S}$ to the kinetic temperature $T_{\rm K}$, the Ly$\upalpha$ photons also exchange energy with the hydrogen atoms through repeated scattering \citep{Chen_2004, Chuzhoy2007, Ciardi2010, Mittal_lya, Reis_2021, Raste_2024_Lyal}. Ly$\upalpha$ heating is a critical and non-negligible component in shaping the thermal history of IGM, which might compete with X-ray heating \citep{Reis_2021, Mittal_echo}. Despite its importance, this heating term is often overlooked.

These competing heating and cooling processes profoundly shape the observable redshifted 21-cm signal. The 21-cm signal (Fig.~\ref{fig:lightcone-global-ps}) traces the contrast between the IGM temperature and that of the radio background radiation (often assumed to be the CMB). As soon as the considerable population of first heating sources is formed, it acts to reduce the contrast between the IGM and the CMB temperatures, eventually heating the IGM above the temperature of the CMB. This action is directly visible in the global 21-cm signal (top panel, Fig.~\ref{fig:lightcone-global-ps}), which transitions from a deep absorption trough into an emission bump \citep[although if heating is inefficient, the signal remains in absorption][]{Fialkov_2014}.  

The spatial fluctuations in the IGM temperature and ionization state, driven by these heating mechanisms, are imprinted on the 21-cm signal (bottom panel, Fig.~\ref{fig:lightcone-global-ps}). For instance, inhomogeneous X-ray heating creates large-scale temperature fluctuations that can dominate the 21-cm power spectrum at intermediate redshifts $z\sim10-20$  (middle panel, Fig.~\ref{fig:lightcone-global-ps}), offering a unique probe of the sources and efficiency of early universe heating \citep{Pritchard_2007}. 
Understanding and modelling these heating processes is thus paramount for interpreting the 21-cm observations with the SKA. 



\subsection{Ionization process}
\label{sec:ion_process}
The reionization process is complex, driven by a variety of ionizing sources. The photon production was likely dominated by softer-spectrum stellar sources, likely with some contribution from harder ones like QSOs and X-ray binaries. 
It has now been established that the process is inside-out, i.e. denser regions are reionized on average earlier, with reionization starting from the high-density peaks, where the early galaxies form, and gradually expanding outwards, with the largest, deepest, emptiest voids reionizing last \citep{2004ApJ...609..474B, furlanetto2004growth, 2006MNRAS.369.1625I, 2024MNRAS.533.2364G, 2025MNRAS.537.2273C}. Furthermore, according to the statistics of the peaks of a Gaussian field, early, rare sources tend to cluster together. Low-mass sources could be suppressed by radiative feedback, thereby boosting the importance of the larger, more clustered sources. Finally, the effect of absorbers, specifically Lyman-limit systems 
\citep{2016MNRAS.458..135S} and 
unresolved gas density fluctuations that increase the recombination rate, is to slow and extend reionization, while also breaking up the 
ionized patches into 
smaller ones
\citep{2021MNRAS.504.2443B,2025arXiv250403384C}.  
Therefore, 
the reionization patchiness encodes a wealth of information about the nature, abundance and clustering of the early galaxies and the properties of the IGM.  

The ionizing sources and sinks, and the resulting reionization history and morphology are reflected in 
the redshifted 21-cm signal and can thus be constrained by it. The detailed dependencies are complex, and subject of ongoing studies, but certain general trends have become clearer. More efficient feedback and suppression of the lower-mass sources increase the relative contribution of the massive sources. The latter are rarer and more clustered, 
yielding larger, smoother
ionized patches. This 
boosts the large-scale 21-cm power and moves the peak of the 21-cm rms fluctuations to larger scales and lower redshifts, making the detection of these signals easier since SKA is more sensitive there, while foregrounds are weaker at late times. In contrast, a less efficient feedback would increase the contribution of the much more numerous low-mass sources, thereby extending the reionization process and producing larger number of smaller ionized patches, making detection more difficult \citep{2012MNRAS.423.2222I}. The same is true also for the tracking of the mean reionization history
(see Fig.~\ref{fig:lightcone-global-ps} for an example), where a shorter and sharper `step' would significantly facilitate detection. Other effects like the spectra of the dominant source population
\citep[e.g.][]{Gessey_2023,Ma2023MNRAS}, metal enrichment and the transition from metal-free to later generations of stars, and even source radiation anisotropy \citep{2025arXiv250502716S} have also been considered.

Many 
redshifted 21-cm line signatures have been investigated over the last 30 years, from the global mean reionization history \citep[e.g.][]{2019MNRAS.483L.109M}, through statistical measures like the power spectra, 1-point  (PDFs, rms, skewness) and 2-point (bispectra) statistics \citep{2006MNRAS.369.1625I,giri2019position, Abinash_nonGaussian2019,2024JCAP...10..003N, munoz21}, line-of sight effects 
\citep[e.g.][]{2013MNRAS.435..460J, Rajesh_lightcone2020, ross2021redshift, thelie25}, 
to direct imaging \citep[e.g.][]{giri2018optimal,giri2019neutral,zackrisson2020bubble}.    
The ionizing photon sinks can also have considerable effect on the 21-cm signatures. During EoR the ionizing photons typically do not travel very far from their emitting galaxy and are absorbed by the nearest neutral patch. However, towards and after overlap, such patches become rare and the mean free path of the ionizing photons rises dramatically \citep{2022MNRAS.516.3389L,2024MNRAS.533..676S},  set by the abundance and nature of the remaining absorbers, Lyman-limit and  Damped Ly$\upalpha$ systems. Furthermore, most of the neutral hydrogen at that epoch is found in the same absorber systems. Consequently, they directly influence the 21-cm power spectra and other related observables and should be included in any realistic modelling, particularly of the tail-end of reionization and post-reionization IGM \citep{2016MNRAS.458..135S,2021MNRAS.504.2443B,2024MNRAS.533.2364G,2025ApJ...979..150F,2025MNRAS.536.3689G}. 









The links between the early galaxies and the 21-cm signal are complex, and thus linking them reliably requires realistic modeling. 
Models range from the cheaper and faster, but inevitably more approximate semi-numerical approaches \citep[e.g.][]{Fialkov_2014}, through simplified simulations \citep{2018MNRAS.476.1741G}, to full and thus more expensive radiative transfer \citep{Mellema06,2024A&C....4800861H} and fully coupled radiative hydrodynamics numerical simulations \citep{2020MNRAS.496.4087O}. Machine learning methods are used to construct fast emulators and thus to further accelerate parameter searches and fitting \citep[e.g.][]{ghara2020constraining}. Currently the parameter space searches and fitting is largely limited to the fast, semi-numerical approaches and  simplified simulations, and emulators based on such modelling, but with the advent of GPUs and improved numerical methods even full radiative transfer methods are becoming viable for this \citep{2024A&C....4800861H}.






\subsection{Accurate 21-cm line transfer for SKA-Low tomographic studies}
SKA-Low's 21-cm tomographic data are light-cone measurements. An observed frequency picks up photons that redshifted into that band after being emitted at specific cosmic time. Each channel is therefore a path-integrated record of the cosmological medium they traversed. As radiation travels through the expanding, inhomogeneous Universe, it undergoes emission and absorption, and interacts with ambient radiation fields, 
such as the CMB radiation 
and any background and ambient 
radiation in the bands. 
The CMB is not merely a continuum 
that the 21-cm signal is observed against. 
It contributes to setting 
the radiation temperature, $T_{\rm r}$, 
and it regulates absorption and emission though radiative (stimulated) transitions. 
Additional radio background would alter this balance. The cosmological 21-cm line radiative-transfer is not simply a convolution. The measured signal encodes the cumulative, frequency-dependent imprints of hydrogen 
density, gas velocities, and radiation fields along its path. As such, 
the radio transfer of the 21-cm line  
is required to be 
self-consistent in the joint 
$(z,\nu)$ domain. 
A resolution is 
  to adopt the cosmological 21-cm line radiative transfer (C21LRT) formulation. 
The corresponding C21LRT equation 
  for  the intensity of the 21-cm line, $I_{L,\nu}$, 
  along the photon path $s$ 
  would take the form: 
\begin{equation}
\label{eq:covariantlineRTstiemiInc} 
\frac{\rm d}{{\rm d} z} \left(\frac{I_{{\rm L},\nu}}{\nu^3}\right)
   = (1+z)\;\!
    \Bigg[
    -\left(\kappa_{{\rm C},\nu}+ \kappa_{{\rm L}, \nu}\, \phi_{\nu}  \,[1-\Xi]\;\!
    \right)  
      \;\!\left(\frac{I_{{\rm L}, \nu}}{\nu^3}\right) 
       +  \frac{(\epsilon_{{\rm C}, \nu} 
     +\epsilon_{{\rm L}, \nu}\,\phi_{\nu})  }{\nu^3}\;\! 
     \Bigg] 
     \frac{{\rm d}s}{{\rm d}z}  \ ,
\end{equation}
\citep{2024MNRAS.531..434C, 2024MNRAS.531.3088W}, 
with the subscript ``L'' denoting the line centre and “C” the continuum underneath and neighbouring to the line. The factor $[1-\Xi]$ accounts for stimulated emission. 
The increment of path length with respect to the change in redshift, ${{\rm d}s}/{{\rm d}z}$, is set by the assumed 
  cosmological model. 
The line absorption and emission coefficients ($\kappa_{{\rm L}, \nu}$ and $\epsilon_{{\rm L}, \nu}$) 
are determined by the Einstein coefficients, 
and 
the number densities of HI in the two hyperfine states. 

The observed 21-cm spectral signal can be expressed either as the specific intensity contrast,
  $(I_{{\rm L},\nu}-I_{{\rm C},\nu}),$  
or as the differential brightness temperature,
  $\delta T_{\rm b} = (I_{{\rm L},\nu}-I_{{\rm C},\nu})(c/\nu)^2/(2k_{\rm B})$. 
This general expression makes 
  no assumptions regarding the optical depth, spatial uniformity of the neutral hydrogen gas, nor linearity in the velocity field.

The accurate modeling of 21-cm tomographic signals that interface simulated fields like ionized fraction, gas density, velocity and temperature with C21LRT to produce light-cone spectral cubes. These are then propagated through the SKA measurement equations to generate 21-cm model templates. 
This modeling approach 
  will enable more proper and 
  accurate comparisons 
  to observational data. 
It will also provide data test-beds 
 to deliver better and more reliable inference of reionization history, bubble size distributions, and the thermal evolution of the gas.


\section{SKA observational setup}\label{sec:SKALow}
We assess the capability of SKA-Low to measure key cosmological and astrophysical signatures by evaluating and comparing its proposed AA* and AA4 antenna layouts. Our analysis considers three primary sources of contamination:
\begin{itemize}
    \item Cosmic variance: This statistical uncertainty is dictated by the finite volume of the observable universe, which is limited by the size of the primary beam. The beam size can be approximated as $21\mathrm{cm}(1+z)/D$, where $D$ represents the diameter of an individual antenna station. For this study, we adopt a station diameter of 38 m.
    \item Instrumental noise: This source of error is influenced by the telescope's design, including antenna sensitivity and array layout, as well as specific observational parameters. See, e.g., \citet{mellema2013reionization} and \citet{giri2018optimal} for a more detailed discussion on these parameters. We evaluate the AA* configuration (307 stations) and the AA4 configuration (512 stations). For both layouts, we adopt a fiducial setup with a total integration time of 1080 hours, a bandwidth of 10 MHz, and a power spectrum binning of $\Delta \ln k = 0.5$.
    \item Foreground Contamination: Bright astrophysical foregrounds represent the most significant challenge. A common strategy to mitigate their impact on the 21-cm power spectrum is to apply a scale cut, removing the Fourier modes most affected by foreground leakage. We adopt a conservative cut at $k \lesssim 0.15$ Mpc$^{-1}$ \citep{Pober_2014}. We should note that some signatures, such as the velocity acoustic oscillations (Sec.~\ref{sec:vao}), are found at larger scale and would require sophisticated foreground mitigation. For a more comprehensive discussion on foregrounds, we refer the interested reader to the \citet{SKAforegrounds}.
\end{itemize}
These mock observations can be generated using publicly available software packages that incorporate the latest SKA layouts, such as \textsc{21cmSense}\footnote{\url{https://github.com/rasg-affiliates/21cmSense}} \citep{murray202421cmsense}, \textsc{Tools21cm}\footnote{\url{https://github.com/sambit-giri/tools21cm}} \citep{giri2020tools21cm}, and \textsc{Karabo}\footnote{\url{https://github.com/i4Ds/Karabo-Pipeline}} \citep{sharma2025karabo}. 



\section{Standard Signatures}\label{sec:StandardSignatures}
Here we summarize the main redshifted 21-cm signatures expected within the $\Lambda$CDM framework.

\subsection{Early galaxies and their properties}\label{sec:early-galaxies}
Early galaxies, formed within the first billion years of cosmic history, critically influenced the thermal and ionization states of the IGM during the Cosmic Dawn and the EoR (e.g. \citealt{Bouwens2015ApJ,Munoz2022MNRAS}). Characterized predominantly by low metallicities ($Z \ll Z_\odot$) and high specific star formation rates (sSFR) \citep{Ceverino2018MNRAS}, these galaxies hosted massive, short-lived stellar populations emitting copious amounts of X-ray, Ly$\upalpha$ and UV ionizing photons. The interaction of these photons with neutral hydrogen in their surroundings left distinctive imprints observable as redshifted 21-cm emission and absorption signals, providing unique probes into the astrophysical processes operating during these epochs.

Establishing robust links between detailed properties of primordial galaxies and observable 21-cm signatures remains an essential objective in contemporary cosmological research. In this regard, the SKA aims to achieve unprecedented sensitivity and spatial resolution to measure the 21-cm fluctuations across a wide range of redshifts. This observational capability will enable stringent tests of galaxy formation and evolution models, and deeper insights into the feedback processes operating between early galaxies and the IGM. 
Particularly, in regular EoR galaxy surveys, only the ``tip of the iceberg'' (the few brightest ones among all the population) has been observed. The majority of the EoR galaxy population is below the detection limit and their properties are largely unknown \citep{Yue2016MNRAS,Yue2018ApJ}. In contrast, the SKA observations  will allow us to reveal the properties of these faint galaxies, as they dominate the ionizing photons budget \citep{Santos2011A&A,Qin2021MNRAS}.

\begin{figure*}
    \centering
\includegraphics[width=0.8\linewidth,trim=0.0cm 0.3cm 0.0cm 0.2cm, clip]{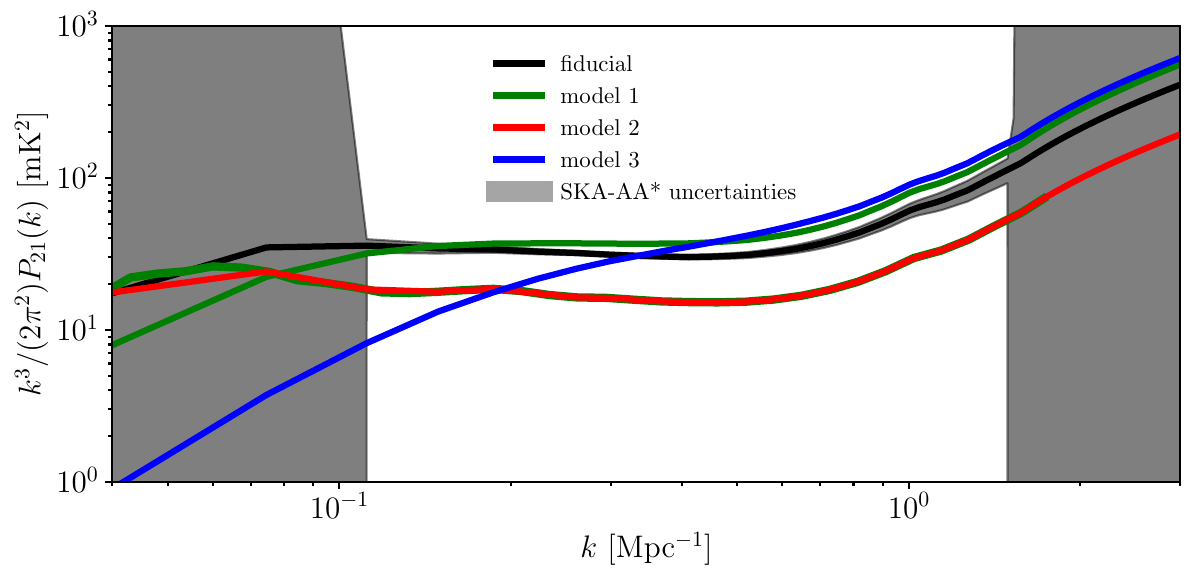} 
    \caption{The 21-cm power spectrum at $z\sim 8$ for different parameters in $f_\star-M_{\mathrm{h}}$, $f_{\rm duty}-M_{\mathrm{h}}$ and $f_{\rm esc}-M_{\mathrm{h}}$ relations. 
    Fiducial model: $\epsilon_0=0.13$, $\gamma_{\rm lo}=0.46$, $\gamma_{\rm hi}=0.82$, $M_p=2\times 10^{12}~M_\odot$, $M_{\rm turn}=5\times 10^8~M_\odot$, $M_{\rm esc,0}=10^{10}~M_\odot$, $f_{\rm esc,0}=0.1$ and $\alpha_{\rm esc}=0.5$. Model 1: $M_{\rm turn}=2\times 10^8~M_\odot$ and $M_{\rm esc,0}=3.2\times 10^9~M_\odot$, others same to fiducial model. Model 2: $\epsilon_0=0.02$ and $\gamma_{\rm lo}=0.17$, others same to fiducial model. Model 3: $\epsilon_0=0.25$, $\gamma_{\rm lo}=0.70$ and $M_{\rm turn}=2\times 10^8~M_\odot$, others same to fiducial model.
    The filled region is the SKA-low AA*  uncertainties with $t_{\rm obs}=1080$ hr and a bandwidth of 10 MHz, calculated by using {\tt 21cmSense}, assuming moderate foreground removal model. 
    }
    \label{fig:early-galaxies-21cmPS}
\end{figure*}

The observable features of the 21-cm signal
are intricately tied to galaxy-specific parameters \citep{Greig:2015qca}. The primary ones are star formation efficiency $f_\star$ and escape fraction of ionizing radiation $f_\mathrm{esc}$. 
Motivated by matching the observed high-$z$ UV LFs by HST and JWST, the star formation efficiency $f_\star$ can be parameterized as \citep{Yang_2003MNRAS}
\begin{equation}
f_\star(M_{\mathrm{h}})=\frac{2\epsilon_0}{\left(M_{\mathrm{h}}/M_{\mathrm{p}}\right)^{-\gamma_{\rm lo}}+ \left(M_{\mathrm{h}}/M_{\mathrm{p}}\right)^{\gamma_{\rm hi}} }.
\end{equation}
This formalism, though very simple, reveals an essential fact that star formation is sensitive to feedback: for halos with $M_{\mathrm{h}} \ll M_{\mathrm{p}}$ the supernova feedback dominates, while for halos with $M_{\mathrm{h}} \gg M_{\mathrm{p}}$ the AGN feedback dominates. Moreover, star formation mode in halos close to $T_{\rm vir}\sim 10^4\,$K is bursty, at a given redshift snapshot, only a fraction of halos host active star formation. This is modelled as a duty cycle \citep{Park_2019MNRAS}
\begin{equation}
f_{\rm duty}(M_{\mathrm{h}})=\exp\left(- M_{\rm turn}/M_{\mathrm{h}}\right).
\end{equation}
When $M_{\mathrm{h}}$ increases to $\sim 10^8-10^9~M_\odot$, $f_{\rm duty}$ approaches to unity (e.g. \citealt{Jaacks_2012MNRAS,OShea_2015ApJ}). The escape fraction of ionizing photons, by contrast, is higher for smaller halos, also partially due to supernova feedback. A parameterization is \citep{Park_2019MNRAS}  
\begin{equation}
f_{\rm esc}=f_{\rm esc,0}\left(M_{\mathrm{h}}/M_{\rm esc,0} \right)^{-\alpha_{\rm esc}},~f_{\rm esc} \le 1.
\end{equation}
The variations of above parameters affect the spatial  characteristics of the 21-cm brightness temperature fluctuations, e.g. the power spectrum. Fig. \ref{fig:early-galaxies-21cmPS} shows the 21-cm power spectra for various parameters in the relations, compared with the SKA-AA* observational uncertainties. Inference of parameters from SKA observed 21-cm power spectrum will build the $f_\star - M_{\mathrm{h}}$, $f_{\rm duty}-M_{\mathrm{h}}$ and $f_{\rm esc}-M_{\mathrm{h}}$ relations, and further derive the information of metallicity and initial mass function of stellar populations in early galaxies. 


Deeper interpretation of the SKA observations will require to develop semi-analytical models (SAMs) that are more sophisticated than above phenomenological model (e.g. \citealt{Ma2023MNRAS}).  Since the relations between feedback mechanisms in SAMs and the 21-cm signal are rather complicated, machine learning would be helpful for bridging them and extract advanced information from SKA observed 21-cm power spectrum.

\subsection{Pop III stars in minihaloes}
\label{sec:firststars}
The emergence of the first generation of stars, known as Population III (Pop III) stars
\citep{1971A&A....13..190L,1984ApJ...277..445C, Bromm2013}, represents a pivotal point in cosmic history. These stars are theorized to have formed from pristine, metal-free gas via molecular hydrogen cooling \citep{Haiman1996}, in stark contrast to the metal-enriched stellar populations observed today \citep{2001ApJ...550..890B,Bromm2004, Ventura2025}. Due to the different cooling mechanisms available in the early Universe, Pop III stars are expected to have been predominantly massive ($>10\,$M$_\odot$) and short-lived \citep{Klessen2023}, although some simulations find lower-mass Pop III stars when including turbulence, magnetic fields and ionizing radiation feedback \citep{Sharda2025a, Sharda2025b}. 

The first stars and the subsequent population of X-ray binaries had a profound impact on the Universe and the observable 21-cm signal, leaving potentially detectable signatures \citep{1984ApJ...277..445C, Mirocha2018, Schauer2019, Mebane2020, qin20,  Gessey2022, ventura23, Pochinda2024, Cruz2025, Gessey2025, Hegde2025, Katz2025, Liu2025,  ventura25, Wasserman2025}. Pop III stars were the first sources of light enabling the WF coupling for the first time (subsection \ref{sec:Lyacoupling}), thus making the cosmic dawn 21-cm signal observable against the CMB. The remnants of these massive Pop III stars, particularly the luminous high mass X-ray binaries \citep{Sartorio2023}, are predicted to have played a crucial role in the evolution of the 21-cm signal by ushering the Universe into the era of cosmic heating (subsection \ref{sec:heating}). Furthermore, the massive Pop III stars are efficient in producing ionizing radiation \citep[e.g.][]{Schaerer2002, Wasserman2025, Liu2025} and the copious soft-UV emission from the Pop~III stars in the Lyman-Werner bands is thought to destroy the $H_2$ molecules required to cool the pristine primordial gas, thus locally delaying or even inhibiting star formation \citep{Haiman1996, 2012ApJ...756L..16A} and resulting in enhanced fluctuations in the 21-cm signal \citep{Holzbauer2012, Fialkov2013}. The first supernovae initiated the process of metal enrichment and the transition to enriched stellar populations \citep[][]{Pallottini2014, ventura25}. Another unique signature that can be imprinted in the 21-cm signal by star formation in minihalos is the enhanced acoustic oscillations in the cosmic dawn power spectrum due to the relative motion between dark matter and gas \citep[see section \ref{sec:vao} and][ for more details]{TseliakHirat, Fialkov2013, Zhang2024}.  These fluctuations are modulated by Pop III star formation, can probe small-scale density fluctuations, and are sensitive to the nature of dark matter.

\begin{figure}
        \centering
        \includegraphics[width=\linewidth,,trim=0.0cm 0.3cm 0.0cm 0.2cm, clip]{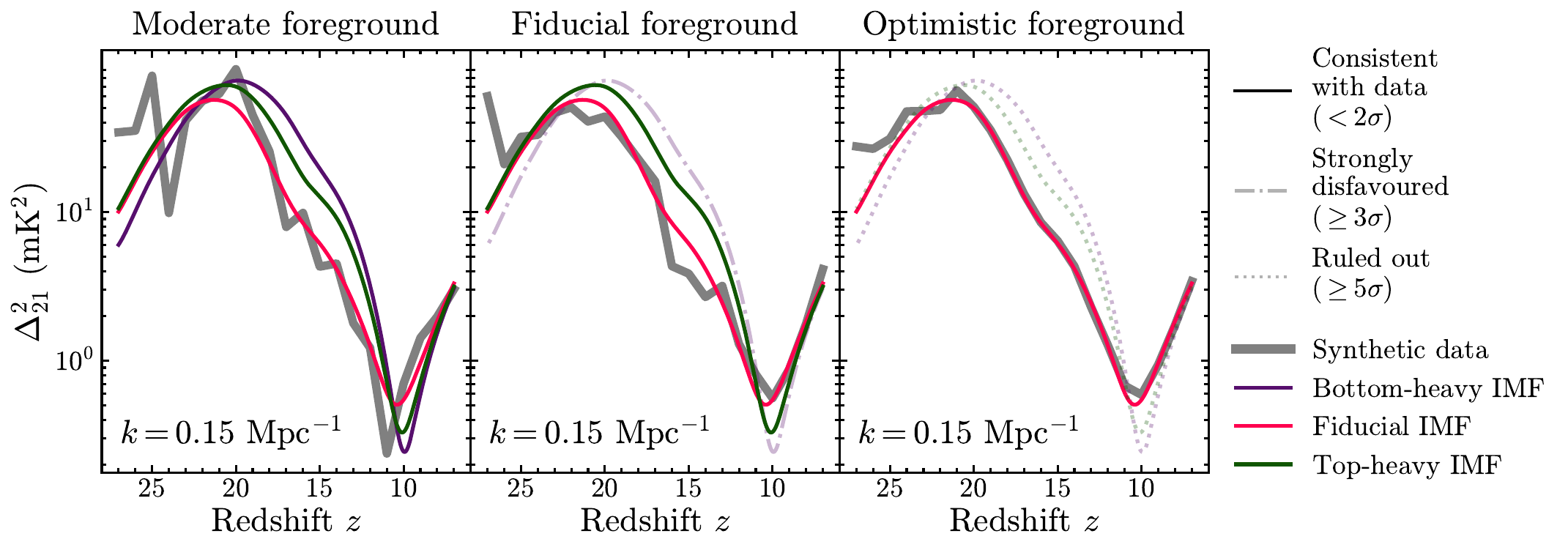}
        \caption{Forecasted Bayesian constraints on the Pop III IMF from the 21-cm power spectrum. We show the 21-cm power spectra simulated using the semi-numerical code \texttt{21cmSPACE} for three candidate truncated power-law Pop III IMFs (different line colours), of the form $dN/dM \propto M^{-\alpha_{\rm III}},\,\, M \in [M_{\rm min}, M_{\rm max}]$. Our bottom-heavy, fiducial and top-heavy IMFs correspond to the \texttt{Sal}, \texttt{Int-1} and \texttt{Top} IMFs considered in \cite{Gessey2025} respectively. The synthetic data are generated assuming a fixed fiducial IMF and noise levels simulated with \texttt{21cmSense} for the full `AA4' SKA antenna configuration with $10$ MHz channel bandwidth and 3000 hours of observation, assuming three different foreground avoidance scenarios: `moderate' (left), `fiducial' (centre) and `optimistic' (right). In the `optimistic' scenario, only k-modes within the main lobe of the beam are excised. In the `fiducial' scenario, we discard all k-modes within the horizon limit, and in the `moderate' scenario an additional buffer of $0.1$ $h\text{Mpc}^{-1}$ above the horizon is removed. The line-style corresponds to the confidence with which each model is ruled out by the data, as summarized in the top right corner of the Figure.}
        \label{fig:ps_IMF_constraints}
    \end{figure}

These signatures of the Pop III stars and their remnants in the 21-cm signal depend on the masses of the first stars \citep[e.g.][]{Mebane2020, Gessey2022, Gessey2025, Liu2025, Wasserman2025}. 
While the global (sky-averaged) 21-cm signal provides insights into the average thermal and ionization history, the true wealth of information about Pop III stars and their mass distribution lies in the spatial fluctuations of the 21-cm signal. Different models for Pop III star formation, their initial mass function (IMF), and their X-ray emission efficiencies are predicted to leave distinct imprints on the 21-cm power spectrum \citep{qin20, Gessey2022,  Gessey2025, Liu2025, Wasserman2025}. The Pop III IMF is strongly linked to the stellar emissivity in the Lyman band \citep{Gessey2022}, the ionizing properties of stars \citep{Schaerer2002, Liu2025} as well as the average lifetime, abundance and luminosity of the subsequent XRB population \citep{Sartorio2023}. Thus, we expect the precise timing and intensity of features in the 21-cm power spectrum to be sensitive to the Pop III IMF across the wide redshift range observable by the SKA \citep[Figure \ref{fig:ps_IMF_constraints} and ][]{Gessey2025}. 

In Figure \ref{fig:ps_IMF_constraints}, we present the 21-cm power spectra at $k=0.15~\rm{Mpc}^{-1}$ for three Pop III IMF types generated using the semi-numerical code \texttt{21cmSPACE}\footnote{To create these curves, we fixed the remaining 21-cm model parameters, including Pop II physics, at the values specified in Table S.1 of \cite{Gessey2025}.}: a bottom-heavy case dominated by small-mass stars, a fiducial IMF, and a top-heavy IMF dominated by massive stars, \citep[corresponding to the \texttt{Sal}, \texttt{Int-1}, and \texttt{Top} IMFs considered in][respectively]{Gessey2025}. The top-heavy case results in more efficient X-ray heating compared to the bottom-heavy case \citep{Sartorio2023}. This discrepancy leads to distinct power spectrum signatures in the 21-cm signal \citep{Gessey2025}, suggesting that models with strong Pop III X-ray emission are distinguishable from models with no or mild Pop III X-ray emission using 1,000 hours of foreground-avoidance observations with the upcoming SKA1-Low. However, as we learn from Figure \ref{fig:ps_IMF_constraints}, robust foreground treatment will likely be necessary to produce 3 -- 5$\sigma$ constraints on the IMF, with foreground avoidance alone being unlikely to yield constraints stronger than $\sim 2\sigma$. In Figure \ref{fig:ps_IMF_constraints}, we show the synthetic 21-cm data created using the fiducial IMF \citep[following the procedure outlined in Section II of][but using updated noise levels simulated with \texttt{21cmSense}]{Gessey2025}. We then calculate the likelihood that the data are generated with a different IMF (bottom-heavy or top-heavy), assuming three foreground avoidance scenarios: moderate, fiducial, and optimistic (see the caption of Figure \ref{fig:ps_IMF_constraints} for details). We find that an aggressive foreground treatment is required to recover the true IMF at $>3\sigma$ within 3,000 hours of observation.



The SKA-Low, with its unprecedented sensitivity across a wide range of redshifts, is uniquely positioned to measure subtle fluctuations driven by feedback from Pop III stars in minihaloes. The ability to simultaneously probe a range of scales and redshifts will be crucial for breaking degeneracies between different models of the first stars and their evolution. This will enable inference of elusive parameters such as the Pop III IMF, in the manner demonstrated with mock data in Fig. \ref{fig:ps_IMF_constraints}.


\subsection{X-ray binaries }
\label{sec:XRBs}
X-ray emission is another key ingredient of the early Universe, contributing to both the heating of the IGM and the ionization of the neutral hydrogen. Much remains uncertain: soft X-rays ($E\sim20\,\mathrm{eV}$) in the form of thermal emission from supernova remnants \citep{2001ApJ...553..499O,2001ApJ...563....1V,2004MNRAS.352..547R} and mini-quasars \citep{2006ApJ...637L...1K,Ciardi2010}, hard X-rays ($E\sim3\,\mathrm{keV}$) from black hole X-ray binaries \citep[XRBs;][]{Mirabel_2011,2013ApJ...764...41F}, and AGN have all been suggested as dominant sources of X-ray emission. The properties of the X-ray background such as its strength, inhomogeneous heating effect \citep{Pritchard_2007,2014MNRAS.445..213F}, and shape of its spectral energy distribution \citep{Fialkov_2014,2014MNRAS.445..213F,Pacucci2014,2017MNRAS.468.3785R} affect the evolution of the IGM temperature, and thus the 21-cm signal coupled to it. The contribution of ancient X-ray sources to the unresolved X-ray background seen today sets an upper limit on their emission efficiency, and is thus a useful observable  \citep{Fialkov_2017, Pochinda2024}. X-rays also contribute to reionization, and can produce a fairly homogeneous low level of ionization. However, since heating gas to the CMB temperature requires far less energy than ionizing it, in most standard models the X-rays dominate early cosmic heating but ultra-violet photons dominate cosmic reionization.

XRBs are thought to be the most likely dominant source of X-ray emissions at $z\gtrsim6$, in particular high-mass X-ray binaries \citep{2013ApJ...776L..31F,2016ApJ...825....7L}. A simple model for their X-ray luminosity, following local starburst-like galaxies, assumes proportionality to the halo star-formation rate \cite[e.g.,][]{2003MNRAS.339..793G,2010ApJ...724..559L,2012MNRAS.419.2095M,2014MNRAS.437.1698M}:
\begin{equation}
L_{\mathrm{X}}/\mathrm{SFR} = f_{\mathrm{X}} \times 3\times 10^{40}{\,\mathrm{erg} \,\mathrm{s}^{-1}}\,\mathrm{M}_{\odot}^{-1}\,\mathrm{yr}\ ,
\end{equation}
where $f_{\mathrm{X}}$ is the X-ray emission efficiency of sources at high-redshifts, normalized to low redshift low metallicity X-ray binaries \citep{2013ApJ...764...41F,2013ApJ...776L..31F,2021ApJ...907...17L,Fialkov_2017}. 

Given the weak current observational constraints on high-redshift X-ray sources, particularly on any temporary high-redshift population that may have had different properties than those of observed sources at lower redshifts, the X-ray efficiency is typically the dominant astrophysical uncertainty in standard astrophysical models of the 21-cm signal. This parameter may vary by up to several orders of magnitude below or above the ``standard'' value of unity, in turn making the predictions of the 21-cm power spectrum vary by several orders of magnitude. Low X-ray efficiency gives late X-ray heating, allowing the early IGM more time to cool adiabatically, producing stronger 21-cm absorption and extending this strong signal to lower redshifts. Thus, among standard astrophysical models, these yield the strongest 21-cm fluctuations and are the first models to be potentially detectable. Models with $f_X=0$ are already strongly disfavoured or ruled out \citep[e.g.,][]{2023ApJ...945..124H}, although this corner of the parameter space is particularly sensitive to weak heating sources, in particular Ly$\upalpha$ heating \citep[e.g.,][]{Chen_2004,Chuzhoy2007, Mittal_lya, Reis_2021} [Note that CMB heating is thought to be negligible \citep{venumadhav18,Meiksin21}]. As long as $f_{\mathrm{X}}$ is not too high, the hard X-ray spectrum expected for XRBs could produce a late heating during cosmic reionization, leading to a variety of signatures that should be observable with the SKA \citep{Fialkov_2014,2014MNRAS.445..213F,Pacucci2014,2017MNRAS.468.3785R,2019MNRAS.482.2653W}. 

In summary, an SKA  detection of the 21-cm power spectrum during cosmic dawn will strongly constrain the intensity and spectrum of X-ray heating sources, at redshifts for which there are almost no other observational constraints. The redshift evolution of X-ray heating will help us determine if it is consistent with XRBs as the primary source.

\subsection{Cosmic rays}
\label{sec:CRs}
While it has been extensively proposed that X-ray photons emitted by the earliest accreting black holes play the dominant role in heating the IGM during cosmic dawn and epoch of reionization, there are alternative heating mechanisms, such as, cosmic ray heating \citep{schlickeiser2002cosmic} that can complement or even compete with traditional X-ray heating, especially during early phases of cosmic dawn \citep{Sazonov_2015, Leite_2017, Jana_2019, Bera_2023, Gessey_2023}. The remnants of energetic supernovae (SNe), which mark the endpoints of massive Population~III stars and Population~II stars, could expand beyond their host dark matter minihalos into the surrounding IGM \citep{2008ApJ...682...49W}, aided by the prior photoevaporation of halo's gas \citep{2004MNRAS.348..753S,2004ApJ...610...14W}, triggered by ultraviolet radiation from the progenitor star. It is argued that, during the evolution of such SN remnants, a considerable fraction of the kinetic energy could have been transferred into cosmic rays, which would eventually escape into the IGM \citep{Hillas_2005, Caprioli_2014}.
Cosmic rays are considered to be generated in the termination shock of the supernova explosions originated from both Population III~and Population~II stars. As a significant fraction of supernovae kinetic energy gets injected into the cosmic rays, they are potential sources of heating of the IGM, and play a significant role in shaping the global 21-cm signal \citep{Bera_2023} and the 21-cm power spectrum \citep{Gessey_2023} during cosmic dawn.
For instance, these sub-relativistic particles are expected to have propagated over large distances and deposited their energy into the IGM, leading to heating by $\sim 10 - 100$ K by redshift $z \sim 15$, prior to the onset of more dominant heating and ionization processes driven by the first galaxies and quasars.
Cosmic rays from young galaxies could also heat and ionize the IGM in the context of both the reionization and post-reionization epochs \citep{Ginzburg_1966, Nath_Biermann_1993, Samui_2005, Samui_2018}. 
Additionally, cosmic rays can originate not only from galaxies but also from microquasars and can have an impact on reionization \citep{Tueros_2014}. 

The low-energy cosmic rays from Pop~III stars can escape their host dark matter minihalos and efficiently deposit energy into the IGM through Coulomb interactions and ionizations \citep{2007MNRAS.382..229S, Sazonov_2015}. In contrast, high-energy cosmic rays from Pop~II stars heat the IGM indirectly through the damping of magnetosonic Alfvén waves, and their effect becomes more prominent at lower redshifts \citep{Samui_2018}.
Depending on the efficiencies of cosmic ray production, the heating can significantly modify the shape and depth of the global 21-cm absorption signal \citep{Bera_2023}. 
Moreover, when combined with scenarios involving dark matter-baryon interactions, which can cool the IGM and produce the unusually deep absorption feature \citep{2018Natur.555...71B, munoz18}, cosmic ray heating can provide sufficient energy to raise the IGM temperature above that of the CMB effectively. The spectral shape of the global 21-cm signal is sensitive to CR parameters, such as acceleration efficiency and spectral slope, offering a new window into constraining early star formation and feedback processes \citep{Bera_2023}. 

Not only the global signal, cosmic rays from early galaxies can significantly affect the fluctuations and morphology of the 21-cm signal \citep[e.g.][]{2018MNRAS.479..153J, Gessey_2023}. The cosmic ray heating introduces a smoother and more spatially extended heating profile compared to soft X-rays, as these particles travel further from their sources before depositing energy. This extended heating leads to a suppression of small-scale power in the 21-cm power spectrum during cosmic dawn, especially at redshifts $z \sim 15-20$. Depending on the energy injection efficiency and propagation properties, the peak of the 21-cm power spectrum may shift to lower redshifts, and its amplitude can be significantly reduced compared to standard X-ray-dominated scenarios \citep{Gessey_2023}. These signatures offer a distinct observable imprint of cosmic ray heating that can be tested with SKA-Low.

On the other hand, in tomographic images of the 21-cm brightness temperature, the presence of cosmic ray heating results in more diffuse and gradual temperature transitions, unlike the sharp contrast between cold and heated regions seen with localized X-ray sources \citep{Gessey_2023}. This smooth temperature distribution can erase the clear bubbles of heating typical of early galaxies, replacing them with large, mildly heated regions. As a result, tomographic maps show a reduced patchiness in brightness temperature and a more uniform heating morphology at high redshifts. Therefore the inclusion of cosmic ray heating not only delays the onset of 21-cm fluctuations but also modifies the spatial structure of the signal, making it a key observable targeting the epoch of reionization and cosmic dawn.

The SKA-Low will enable high-precision studies of cosmic rays in the energy range $\sim10^{16}$--$10^{18}\,\mathrm{eV}$, covering the transition between the Galactic and extragalactic regimes. The dense SKA-Low core, comprising $\sim6\times10^4$ antennas within a $\sim1\,\mathrm{km^2}$ area, will allow detailed radio detection of extensive air showers initiated by high-energy cosmic rays \citep{2017EPJWC.13502003H, 2016Natur.531...70B, SCHRODER20171}. Simulations indicate that SKA can achieve a depth of shower maximum measurement precision of $\sigma(X_{\max}) \simeq 6\,\mathrm{g\,cm^{-2}}$ for $\sim10^{17}\,\mathrm{eV}$ showers, significantly surpassing current radio arrays such as LOFAR \citep{2013A&A...560A..98S}. 

\subsection{Supermassive black holes}\label{sec:SMBHs}
Until now several hundred quasars have been discovered before the end of reionization \citep{2020zndo...3634964B}, among them, one has the highest redshift of $\sim 10.1$ \citep{2024ApJ...960L...1N}, approaching the Cosmic Dawn. These quasars are powered by supermassive black holes (SMBHs) with typical masses $\sim 10^7 - 10^{10} M_\odot$ \citep{2024ApJ...960..112T}. On the other hand, JWST has identified several hundred Little Red Dots (LRDs) above redshift $\sim 5$ \citep{2024ApJ...963..129M}. They are believed to be AGNs powered by SMBHs with masses below $\sim 10^7 M_\odot$ \citep{2025arXiv250316595R}. 
Spectroscopically confirmed SMBHs have a number density $\sim 10^{-4}~{\rm Mpc}^{-3}$  at $z\sim 6$ \citep{2023ApJ...959...39H}, while an estimation based on photometric variability is as high as $\sim 10^{-2}~{\rm Mpc}^{-3}$ \citep{2024ApJ...971L..16H, 2025arXiv250117675C}. 
These high redshift SMBHs have large uncertainties regarding their seeds, number density, and properties, all of which can differ significantly from low redshift ones, and could potentially be constrained from their imprints on the 21-cm signals.
In particular, high redshift SMBHs appear overmassive relative to their hosts, with black hole mass to stellar mass ratios, $M_\bullet/M_*$, of $\sim 0.01 - 0.3$, sometimes approaching unity \--- far above the local value of $\sim 0.001$ \citep{2024Natur.627...59M}. 

The origin of the SMBH seeds and their growth is a long-standing puzzle. Three popular astrophysical scenarios have been proposed \citep{2010A&ARv..18..279V}: light seed, medium seed and heavy seed. The most natural one is the stellar mass black holes ($\sim 10^2 M_\odot$) from Pop III star remnants \citep{2002ApJ...567..532H}, which is the light seed scenario. Runaway collisions of stars in extremely dense stellar clusters may yield heavier seeds ($\sim 10^2-10^3 M_\odot$), and this is the medium seed scenario. Additionally, seeds as massive as $\sim 10^4-10^6 M_\odot$ can form from the collapse of pristine metal free gas in atomic-cooling halos where H$_2$ cooling is suppressed \citep{2006MNRAS.370..289B}, and this is the heavy seed scenario. In this scenario, the gas either collapses into a black hole directly, or experiences a short main sequence phase of a supermassive star with a comparable mass before that. They are all classified as direct collapse black holes (DCBHs).
The  light seed scenario demands nearly continuous Eddington or super-Eddington accretion, which is difficult to sustain \citep{2009ApJ...701L.133A}, while in the medium seed scenario the collapse could only happen under rare, extremely dense conditions. For this reason, DCBHs are  considered to be a promising scenario, as they naturally produce the overmassive black holes with accretion rate close to or even exceeding the Eddington limit \citep{2015MNRAS.448..104P}. Dynamical heating \citep{2019Natur.566...85W} or radiation from other previously-formed DCBHs \citep{2017ApJ...838..111Y} may promote their formation and boost the abundance to above the observed number density level of SMBHs. Moreover, a DCBH has a higher chance to be observed directly, due to its large mass and high accretion rate.

\begin{figure*}
    \centering
\includegraphics[width=0.8\linewidth,trim=0.0cm 0.2cm 0.0cm 0.2cm, clip]{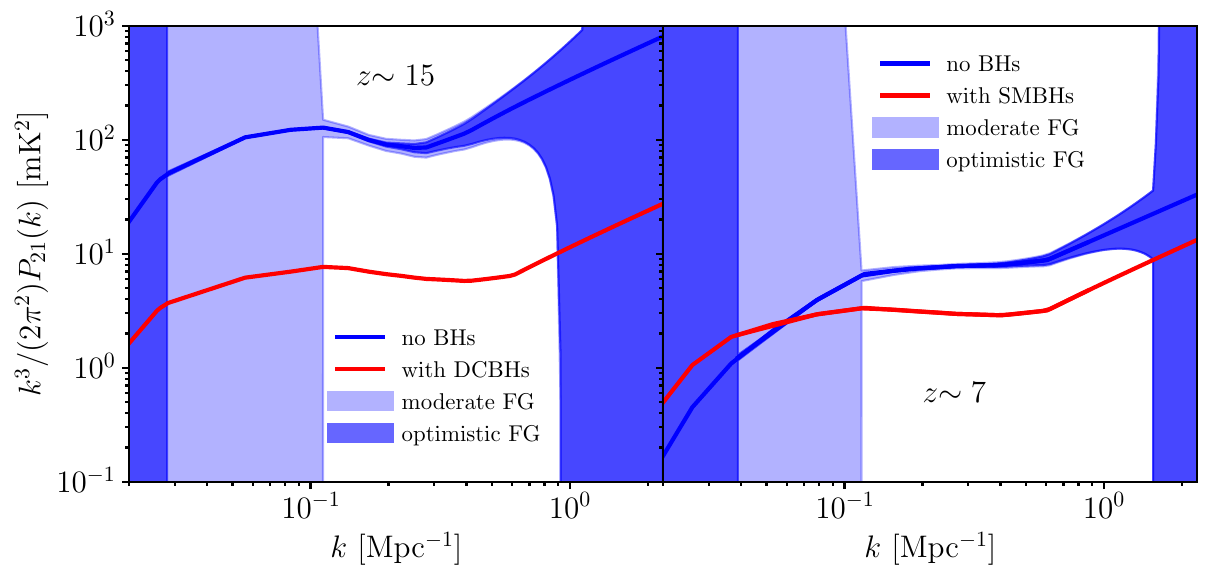}     
    \caption{
     {\it Left:} The 21-cm power spectrum in Cosmic Dawn, in the absence/presence of DCBHs. DCBHs abundance is normalized to $\sim 10^{-2}$ Mpc$^{-3}$ at $z\sim 10$.
     {\it Right:} The 21-cm power spectrum in EoR, in the absence/presence of SMBHs. The SMBHs number density is normalized to be $\sim 2\times 10^{-3}$ Mpc$^{-3}$ at $z\sim6$. In all panels the filled regions refer to the SKA-low AA*  uncertainties with $t_{\rm obs}=1080$ hr and a bandwidth of 10 MHz.  The uncertainties are calculated by using {\tt 21cmSense}, and optimistic (darker) and moderate (lighter) foreground removal models are assumed. 
    }
    \label{fig:BHs-21cmPS}
\end{figure*}

Until now, DCBHs have not been identified in observations. Their number density is constrained by the observed high-z AGNs, since each AGN at least has one seed. Thanks to the special formation process, DCBHs are believed to be Compton-thick at birth, and their ionizing and soft X-ray photons are heavily absorbed. 
Their direct contribution to reionization is negligible, but may still essentially change the thermal history of the IGM and hence the 21-cm signal. The predicted influence on the 21-cm power spectrum is shown in left panel of Fig. \ref{fig:BHs-21cmPS}, in comparison with the uncertainties of the SKA-low AA*.
Note that the uncertainties at the large-scale end are dominated by the residual foregrounds, and the uncertainties at the small-scale end are mainly due to thermal noise and the limited number of long baselines.

In addition to the DCBHs in Cosmic Dawn, the SMBHs will also imprint distinctive features on the 21-cm signals during the EoR.
Compared with regular star-forming galaxies, SMBHs are rarer and more biased, and they can generate giant ionized bubbles around them. Inside these bubbles, the gas is fully ionized and there is no 21-cm signal. At the edge of the ionized bubble, the gas is heated and partially ionized, which produces a strong emission signal.  Well beyond the edge, the 21-cm signal approaches the level of the mean IGM, which is weak emission during the EoR.
According to recent observations carried by JWST, the number density of the high-$z$ SMBHs could be $\gtrsim 10^{-4} {\rm Mpc}^{-3}$ \citep{2023ApJ...959...39H},  and $M_\bullet/M_* \gtrsim 0.01$ \citep{2024Natur.627...59M}. The brighter SMBHs and their giant ionized bubbles are suitable for detection as individual targets. However, based on the SMBHs growth scenario, it is quite possible that there are more fainter SMBHs below the detection limit and/or at higher redshifts, and could be potentially more promising to be detected from their imprints on 21-cm signals. A typical 21-cm observation may have a survey volume $\gtrsim 10^6-10^7 {\rm Mpc}^{-3}$, there must be numerous SMBHs in the volume.
It is challenging to directly image the ionized bubbles around them, however, biased ionized bubbles can influence the statistics of the 21-cm signal, so 21-cm observations can provide information of the SMBHs as well. 

The right panel of Fig. \ref{fig:BHs-21cmPS} show the predicted influence of SMBHs on the 21-cm power spectrum, compared with the observational uncertainties for SKA-low AA*. Indeed, in the presence of SMBHs, the 21-cm signal becomes more anisotropic on large scales, and has a lower power on intermediate scales. The feature on intermediate scales is larger than the uncertainties of SKA-low AA*, so the telescope has the potential to detect the SMBHs via 21-cm power spectrum.
On large scales, the feature is buried in the foreground because if a moderate foreground removal model is adopted. However, in the case of optimistic foreground removal model, the uncertainties would be subdominant, and lager-scale 21-cm power spectrum could also help distinguish features induced by SMBHs.

In summary, DCBHs mainly influence the 21-cm power spectrum at Cosmic Dawn by almost uniformly heating the IGM with hard X-rays, while SMBHs mainly influence the 21-cm power spectrum at the EoR, when the IGM is already heated to $\gg T_{\rm CMB}$, by generating giant biased ionized bubbles.   Moreover, a SMBH also has the chance to be observed directly by the SKA if it is in the radio-loud phase. For a SMBH with mass $\sim 10^8 M_\odot$ and accretion rate 1\% of the Eddington limit, if it is located at redshift 7 and has a radio luminosity following the relation of radio-loud samples in \cite{2022MNRAS.513.4673B}, then its flux would be of the order of 1000 mJy at 200 MHz. This is well above the noise level of SKA-low AA* with 1 MHz bandwidth and 10 hour integration time.

\subsection{Velocity acoustic oscillations}\label{sec:vao}
While regular baryonic matter interacts with radiation, most of the likely candidates for dark matter particles do not. At recombination, this imparts different initial conditions to these two fluids, manifesting as a streaming relative velocity between them.
In \citet{TseliakHirat} it was shown that this relative velocity is supersonic after baryon-photon decoupling (given the relatively low sound speed of baryons post-recombination), and remains so until the baryonic gas is heated by the first luminous sources.
As a consequence, this relative velocity strongly modulates the star-formation rate density, chiefly through three effects: it suppresses the amount of structures formed~\citep{TseliakHirat, naoz12, bovy13}, their gas content~\citep{tseliakhovich11, dalal10, naoz13}, and the star-formation rates in those structures~\citep{greif11, stacy11, oleary12, fialkov12, hirano18, schauer21, hegde23, 2023MNRAS.525.5479C,ventura23, ventura25}.
Regions of large relative velocity will form fewer stars, and in turn their 21-cm signal will be delayed when compared to low-velocity regions, which has now been demonstrated in both semi-numerical models \citep{fialkov12} and full simulations
\citep{2016ApJ...830...68A,2023MNRAS.525.5479C}.

\begin{figure*}
    \centering
\includegraphics[width=0.9\linewidth,,trim=0.0cm 0.5cm 0.0cm 0.4cm, clip]{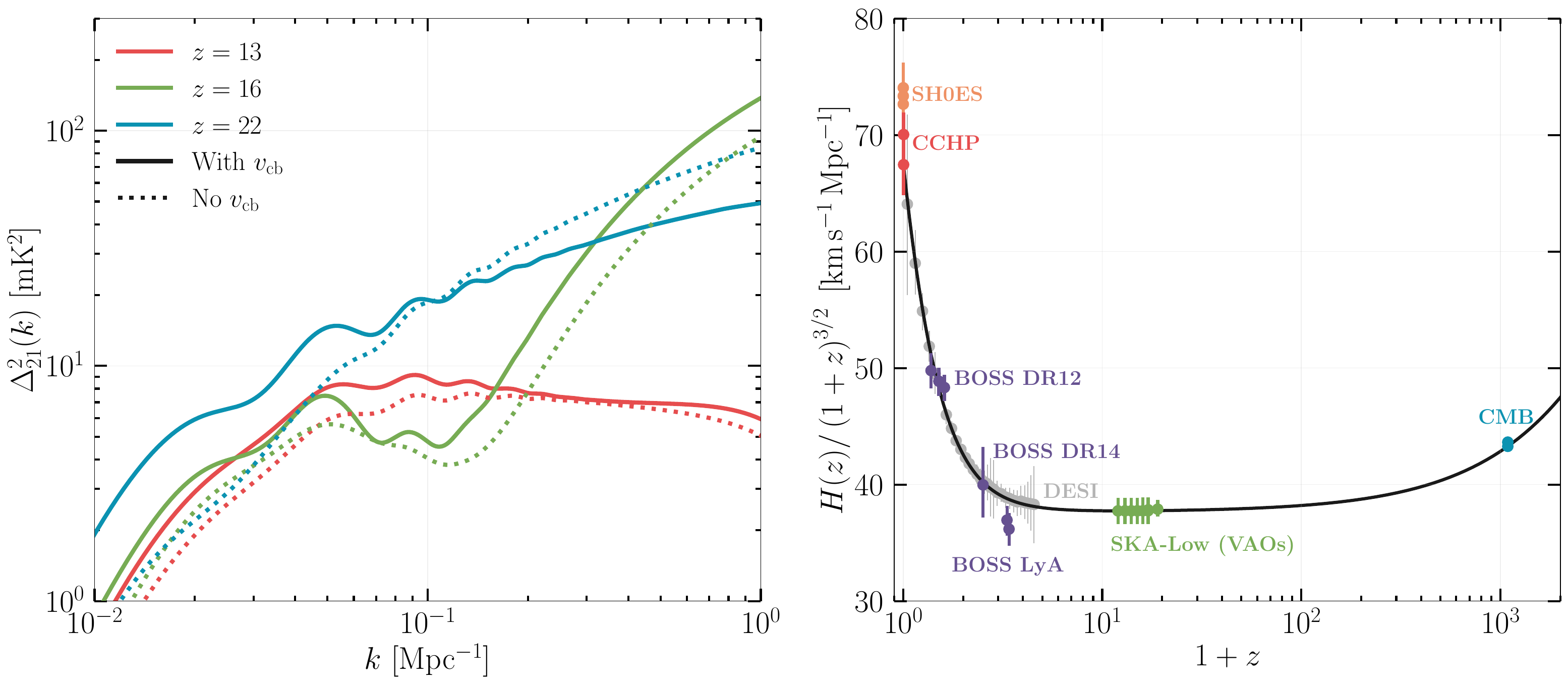}       
    \caption{A depiction of how velocity-induced acoustic oscillations (VAOs) can be used to measure cosmic expansion.
     {\it Left:} The 21-cm power spectrum at $z= 13, 16, 22$ computed with {\tt\string Zeus21}~\citep{munoz2023effective, Cruz2025} in the absence (dotted) and presence (solid) of dark matter-baryon relative velocities $v_\mathrm{cb}$. Including the streaming effects alter the spectral shape, leading to more pronounced acoustic features whose peaks measure the quantity $H(z)r_s$.
     {\it Right:} Existing measurements of the Hubble parameter $H(z)$ across cosmic time with different probes, along with forecasted constraints from future 21-cm experiments. Namely, we plot the constraints from the CMB~\citep{act25, Plankc2018VI} (blue), from forecasts of future \textit{SKA-Low} observations \citep{munoz19b, sarkar23}, galaxy clustering measurements with \textit{BOSS}~\citep{bossdr12, bossdr14, bautista17} (violet) and \textit{DESI}~\citep{desi16, desi24, desi25} (gray), from type Ia supernovae~\citep{riess19, riess22, riess24} (orange), and tRGB measurements~\citep{freedman25, lee25} (red).
    }
    \label{fig:VAOs}
\end{figure*}

The 21-cm maps observed by the SKA will inherit the fluctuations of the dark-matter---baryon velocity imprinted at recombination, which show up as ``wiggles'' on the 21-cm power spectrum~\citep{dalal10, visbal12, Fialkov2013, munoz19b, 2023MNRAS.525.5479C, Cruz2025}.
These wiggles are termed velocity acoustic oscillations (VAOs) and provide a new standard ruler to measure cosmology with the SKA at high redshifts~\citep{munoz19b}.
The VAOs have a different shape than the low-redshift density BAOs (e.g.,~\cite{eisenstein98}), but the same acoustic origin.
Fig.~\ref{fig:VAOs} shows a plot of the 21-cm power spectrum with and without relative velocities at different redshifts. While the power spectra with no streaming velocities exhibit acoustic oscillations on the order of $\sim 5\%$ from the baseline, adding relative velocity fluctuations yields much more pronounced $\mathcal{O}(1)$ oscillations with peaks shifted from the no-$v_\mathrm{cb}$ case. The position of these VAO peaks are sensitive to the cosmic expansion and geometry at previously uncharted cosmic epochs. Their applications include measuring the Hubble expansion rate at $z=10-20$~\citep{munoz19b, sarkar23}, searching for inflationary isocurvature modes~\citep{hotinli21}, testing beyond-concordance cosmological models such as fuzzy dark matter~\citep{sarkar2022exploring, flitter2022closing, hotinli22}, primordial magnetic fields~\citep{2005MNRAS.356..778S, 2006MNRAS.372.1060T, 2008PhRvD..78h3005S, 2009ApJ...692..236S, Bowman_2018, 2019JCAP...01..033K, 2019MNRAS.488.2001M, 2024PhRvD.109b3518C, 2025JCAP...01..089B}, new light degrees of freedom (e.g. light neutrino-like species~\citep{montefalcone25}), and models where dark matter interacts with baryons~\citep{munoz18, sun25, rahimieh25, Liu19}.

Observationally, the SKA is well positioned to observe VAOs.
VAOs always enhance the 21-cm power spectrum, and the presence of relative velocities pushes the signal to lower $z$ (all parameters being equal), both of which aid in detection.
Separating VAOs from the smooth signal, however, requires relatively good $k$ resolution, for which baseline and bandwidth distribution are important.
A promising technique is in the form of AI-inspired techniques such as wedge reconstruction~\citep{kennedy24, sabti25}, which are expected to thrive in the large volumes observed by the SKA. VAOs are most prominent during cosmic dawn when halo masses are lower, and therefore are more susceptible to the suppressive effects of the velocity differential. However, it may be possible to observe VAOs and its relics towards lower redshifts. Small-scale clumping in the IGM eventually become important photon ``sinks'' during reionization but may be subject to pressure smoothing effects from the streaming velocity. Such an effect could impact the ionizing efficiency of early galaxies and the large-scale variation in reionization history~\citep{cain20, park21}.

\subsection{Other standard signatures}\label{sec:other}

In this section we briefly summarize a few other 21-cm signatures that are expected in the standard $\Lambda$CDM model: small-scale clumping, line-of-sight anisotropy, and 21-cm polarization.

In standard CDM, small-scale clumping is predicted to
affect the 21-cm signal. At high redshifts, when star formation is
still rare, dense dark matter clumps form due to structure formation
and pull in gas that is then heated. Due to the non-linearity of these
fluctuations, they do not average out, but change the 21-cm signal as
seen on much larger scales \citep{2006NewAR..50..179A, 2006ApJ...646..681S,2014PhRvD..89h3506A, 2018ApJ...869...42X, cain20,2021MNRAS.504.2443B,2021ApJ...923...98X,2025NatAs.tmp..183P}. In particular,
clumping has a strong impact on the large-scale 21-cm power
spectrum \citep{2021MNRAS.504.2443B,2025arXiv250911175S}; after the Dark Ages, once
Ly$\upalpha$ coupling kicks in due to the first stars, the 21-cm
signal strengthens, and clumping suppresses the observable power
spectrum around redshift 20 by as much as a factor of two. As the
clumping effect arises from small-scale density fluctuations (typical
mass scale of 20 million solar masses), it offers a unique opportunity
to probe the standard cold dark matter model in a new regime and thus
potentially investigate the properties of dark matter, adding new
constraint to models such as warm dark matter and fuzzy dark matter
\citep{fuzzy}. The clumping effect in CDM is significantly higher than
the sensitivity of the planned SKA AA$^\star$ configuration, by up to
a factor of 20, though detection will require separation from
foregrounds and from astrophysical contributions to the 21-cm power
spectrum. It may also be possible to constrain similar scales through their effect on the formation of small halos and the 21-cm effect of astrophysical radiation from those halos \citep{Munoz:2019hjh}.

While most 21-cm analyses in anticipation of the SKA focus on the spherically-averaged 21-cm power spectrum, a significant line-of-sight anisotropy is expected in the standard case, potentially enabling more information to be extracted. Gas motions along the line of sight produce an anisotropy due to the line-of-sight velocity gradient \citep{2005ApJ...624L..65B} that arises through the Kaiser effect of redshift-space clustering \citep{1987MNRAS.227....1K,2004MNRAS.352..142B}. For linear fluctuations, this dependence on the angle to the line of sight makes it possible to measure three separate coefficients \citep{2005ApJ...624L..65B}, corresponding to the power spectrum of density fluctuations, the power spectrum of isotropic 21-cm fluctuations, and their cross-correlation. Numerical investigations during cosmic reionization
\citep{2006ApJ...653..815M,2012MNRAS.422..926M,2013MNRAS.435..460J,2013PhRvL.110o1301S} suggest that the decomposition of the line-of-sight anisotropy is more
complex than the simple linear limit, but it should still provide
additional information that will help track the evolution of 21-cm
fluctuations over various eras \citep{2015PhRvL.114j1303F}. An
additional source of 21-cm anisotropy is the light-cone anisotropy
\citep{2006MNRAS.372L..43B}; the look-back time changes with the
radial distance, and the character of the 21-cm fluctuation sources
evolves with time, which results in a line-of-sight eﬀect that
introduces anisotropy. In particular, this can generate a significant
anisotropy on large scales near the end of reionization, as has been
studied in numerical simulations
\citep{2012MNRAS.424.1877D,2014MNRAS.442.1491D,2014ApJ...789...31L,2014MNRAS.439.1615Z}. Also, if 21-cm data are analyzed using assumed cosmological parameters that
diﬀer from the true ones, this causes an Alcock-Paczynski \citep{1979Natur.281..358A} anisotropy that can be used to constrain
cosmological parameters \citep{2005MNRAS.363..251A,2005MNRAS.364..743N} through an additional anisotropic term added (in the limit of linear fluctuations) to the 21-cm power spectrum \citep{2006MNRAS.372..259B}. Some exotic models can produce a strong line-of-sight anisotropy (e.g., see Sec.~5.1). It remains to be seen to what degree the SKA can measure the signal's intrincsic line-of-sight anisotropy, given that the anisotropy is a fundamental part of algorithms for removing and/or
avoiding the strong foreground.

While most studies of high-redshift 21-cm signatures focus on the total intensity
(Stokes $I$) of the 21-cm signal,
its polarization (Stokes $Q$, $U$ and $V$) presents a complementary window into CD/EoR.
In addition, Faraday rotation of the linear polarization signals offers a possible synergy
between Cosmic Dawn and the cosmic magnetism science goals of SKA \citep{2015aska.confE..92J,2020Galax...8...53H}.
The dominant mechanism for generating the 21-cm polarization signal
is Thomson scattering of the 21-cm photons off free electrons in the ionized IGM \citep{2005ApJ...635....1B}.
Analogous to the CMB polarization, this process converts
the quadrupole moment of the primary 21-cm temperature field into linear polarization,
predominantly of the E-mode type for scalar perturbations.
Therefore, this signal is sourced by a much larger volume
than that for the temperature signal (the latter is restricted to the past light cone),
thus carrying unique information about the distribution and morphology of the ionized regions.
A prediction for this signal was made by \citet{2021ApJ...918...14L},
using semi-numerical simulations that capture inhomogeneous reionization.
The overall amplitude of the angular power spectrum was found to be weaker than the sensitivity of SKA1-Low,
with the peak of the EE spectrum reaching only $\sim1\,\mu$K and the TE spectrum $\sim 0.03\,$mK.
Notably, the TE spectrum exhibits a distinctive zero-crossing at large angular scales ($\ell<100$) that is a potential new diagnostic of reionization physics.
A major challenge to the detection of this E-mode polarization is Faraday rotation from Galactic magnetic fields \citep{2014PhRvD..89l3002D}, but 
cross-correlation approaches may enhance detectability \citep{2023PhRvD.107l3533J}.
Circular polarization of the 21-cm radiation (Stokes $V$)
can arise from the Zeeman effect due to intergalactic magnetic fields \citep{2005MNRAS.359L..47C},
as well as from the interaction between the spin-polarization quadrupole of neutral hydrogen atoms and the quadrupole anisotropy of the CMB,
providing a potential probe of primordial gravitational waves \citep{2018PhRvD..97j3521H,2018PhRvD..97j3522M}.
The predicted signal in $\Lambda$CDM model
peaks at $\ell\sim 400$ and $z\sim 17$ but is still orders of magnitude beyond detection by the SKA \citep{2021PhRvD.103b3516J}.

\section{Exotic Signatures}
\label{sec:ExoticSignatures}

In this section we summarise potential signatures from exotic sources and new physics beyond the $\Lambda$CDM framework.

\subsection{Strongly-emitting radio galaxies}
\label{sec:radiogal}
The Absolute Radiometer for Cosmology, Astrophysics, and Diffuse Emission \citep[ARCADE-2,][]{Fixsen_2011} measurements at GHz frequencies revealed a strong radio background, significantly deviating from the CMB blackbody spectrum at low frequencies. This was corroborated by the first station of the Long Wavelength Array \citep[LWA-1,][]{Dowell_2018} in the $40-80$ MHz range. The amplitude of this radiation exceeds the expectations based on known Galactic and extragalactic radio sources \citep{Singal2018, Singal_2023}, making its origin an interesting astrophysical puzzle. Proposed explanations include exotic  processes such as dark matter annihilation \citep{Fraser:2018, Pospelov:2018}, superconducting cosmic string \citep{Brandenberger:2019}, radiative decay of relic to sterile neutrinos \citep{Chianese2019} in the early Universe or radio emission from accreting supermassive primordial black holes \citep{Mittal_erb}. A more astrophysically grounded explanation of this observed radio excess could be a population of high-redshift radio galaxies \citep{Condon2012}. The Galactic contribution to the radio background is significantly uncertain \citep{Subrahmanyan:2013}, but the observed background certainly puts an upper limit on the contribution from any extragalactic source population. 

The tentative detection of an absorption profile by EDGES \citep{Bowman_2018} at $\sim 78$ MHz (corresponding to a redshift $z \sim 17$) along with the extragalactic radio background measured by ARCADE-2 have spurred significant interest in understanding excess radio emissions from astrophysical sources and their impact on the 21-cm signal. The enhancement of the CMB by an excess radio background, as an explanation for the EDGES experiment, was initially explored without focusing on particular sources \citep{Bowman_2018,feng18,fialkov19}. However, high-redshift galaxies are natural candidates for such a background

To investigate this possibility, \citet{mirocha19} modelled the radio emissivity as proportional to the star formation rate (SFR), as observed in present-day galaxies \citep{gurkan18}, and found that high-redshift galaxies would need to be approximately 1,000 times more radio-luminous than their present-day counterparts (relative to the SFR) to explain the EDGES signal (see also \citet{Mittal_jwst}). We note that the SARAS measurement has presented counter-evidence to the EDGES results \citep{SARAS3}, while other analysis methods disfavour an excess radio background in relation to EDGES \citep{2025A&A...698A.152C}. 

High-redshift galaxies should produce an inhomogeneous radio background due to their clustered distribution. Incorporating this into a 21-cm semi-numerical code, \citet{Reis2020} showed that this inhomogeneity introduces a new type of fluctuations in the 21-cm signal. 
Since the 21-cm absorption occurs along the line of sight, it is therefore sensitive to radio sources lying behind each absorbing cloud. Including this line-of-sight effect \citep{Sikder2023} enhances the 21-cm power spectrum during cosmic dawn, by up to two orders of magnitude. This also induces a new anisotropy in the 21-cm power spectrum. Note that in these models, during the late stages of reionization the radio fluctuations disappear due to the rising abundance of radio sources.  

As noted above, the intensity of the observed extragalactic radio background puts an upper limit on the excess radio emission from high-redshift sources. Additional constraints are placed by the observational upper limits on the clustering of the radio background. In particular, there are measured limits on arcminute-scale anisotropy from the Very Large Array at 4.9 GHz \citep{Fomalont1988} and the Australia Telescope Compact Array at 8.7 GHz \citep{Subrahmanyan2000}. These provide the strongest current constraints on radio emission by high-redshift galaxies \citep{holder2014,Sikder2024_RadioClustering}, although models are still allowed that are consistent with EDGES. 

\begin{figure*}
    \centering
\includegraphics[width=0.45\linewidth,,trim=0.0cm 0.4cm 0.0cm 0.4cm, clip]{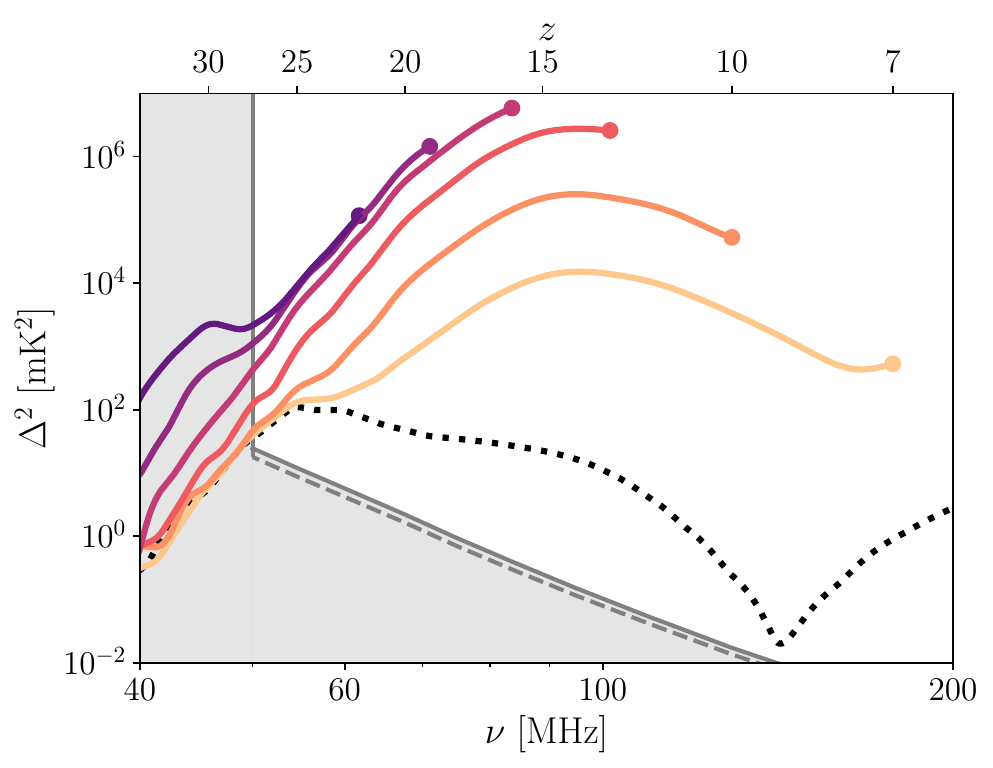}    
\includegraphics[width=0.45\linewidth,,trim=0.0cm 0.4cm 0.0cm 0.4cm, clip]{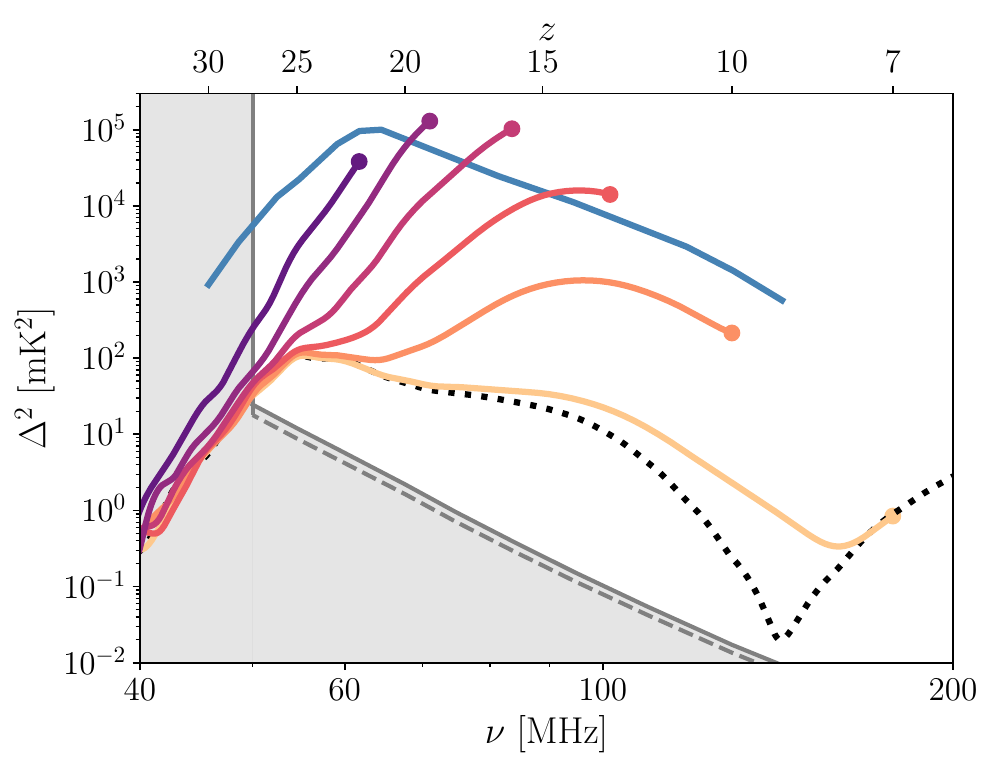}         
    \caption{
    An estimate of the maximum 21-cm power spectrum from models with strongly-emitting radio galaxies, given current observational constraints (based on Fig.~5 of \citet{Sikder2024_RadioClustering}). We show the power spectrum (in terms of the squared fluctuation in mK$^2$) at wavenumber $k=0.1$~Mpc$^{-1}$, versus observed frequency $\nu$ and redshift $z$. Each coloured curve that ends with a dot shows the prediction of a particular model with a fixed radio-emission efficiency, compared to observational constraints assuming that the corresponding galaxy population persisted down to the redshift where the curve ends. {\bf The upper envelope of all these curves should be seen as the prediction.} The ending redshifts are 7, 10, 13, 16, 19, and 22. We also show for comparison the case with no excess radio background, i.e., with the CMB as the assumed background (dotted black line). Also shown is the $k= 0.1$ Mpc$^{-1}$ sensitivity for the SKA configuration AA$^\star$ (solid grey line and the corresponding shaded area) and configuration AA4 (dashed grey line); this shows thermal noise, assuming 1080 hours of total integration time and optimistic foreground removal, using the code 21cmSense  \citep{Pober2013aj,Pober:2014apj}.
     {\it Left:} Includes the constraint on the intensity of the observed extragalactic radio background.
     {\it Right:} Includes also the observational upper limits on the clustering of the radio background. Separately, we show (solid blue curve) an estimate of the maximum power spectrum from a model with a baryon -- dark matter interaction (Sec.~5.3), from Fig.~1 of \citet{imDM}, for the interacting millicharged dark matter model at wavenumber $k=0.13$~Mpc$^{-1}$. 
    }
    \label{fig:RadioDM}
\end{figure*}

Figure~\ref{fig:RadioDM} shows the maximum 21-cm power spectrum from models with strongly-emitting radio galaxies, given current observational constraints \citep{Sikder2024_RadioClustering} on radio galaxy populations from various redshifts. Populations with high radio efficiencies may have existed only at high redshifts. The maximum power spectrum is $6 \times 10^6$~mK$^2$ (from $z=16$) given the constraint on the intensity of the observed extragalactic radio background, lowered to $1.3 \times 10^5$~mK$^2$ (from $z=19$) once we include the observational upper limits on the clustering of the radio background. The latter is still more than 3 orders of magnitude higher than the power spectrum of the same model in its standard version, without an excess radio background. It is 5 orders of magnitude above the expected SKA sensitivity.


\subsection{Primordial magnetic fields}
\label{sec:PMFs}
Magnetic fields exist everywhere in the universe \citep{2004IJMPD..13.1549G, 2013A&ARv..21...62D}. The origin of these cosmic magnetic fields is under discussion, and a possible scenario is that a tiny magnetic field is generated in the early universe, which is called ``Primordial magnetic fields (PMFs)'', and they evolved to be the galactic and/or intergalactic magnetic fields (for a recent review of the PMF scenario, for instance, see \citealt{2016RPPh...79g6901S}).
Note that the astrophysical processes may also produce large-scale magnetic fields in the late universe \citep{1950ZNatA...5...65B, 1976JFM....77..321P, 2005PhR...417....1B}.
In particular, observations of magnetic fields on large scales, such as in void regions or large-scale structure, may invoke generation scenarios in the early universe \citep{2010ApJ...722L..39A, 2010Sci...328...73N}.

PMFs induce small-scale IGM gas perturbations through Lorentz forces on the baryon fluid, in addition to the standard curvature perturbations \citep{1978ApJ...224..337W, 1996ApJ...468...28K}. Unlike standard ones, PMFs induce isocurvature-like fluctuations that are inherently non-Gaussian and strongly scale-dependent. 
These PMF-induced fluctuations can dominate the small-scale power spectrum, especially below the magnetic Jeans scale, and the field strength and scale dependence of PMFs determine the characteristic scale of the induced fluctuations. Consequently, PMFs can amplify structure formation seeds by increasing density fluctuations at high redshifts.

PMFs induce not only density fluctuations, but also the IGM gas temperature fluctuations at high redshifts. There are two main processes by which PMFs heat up IGM gas: ambipolar diffusion and MHD turbulence after recombination. The former is brought by Coulomb scattering acting between neutral and ionized particles in a partially ionized plasma, and the latter is a nonlinear energy cascade process from large-scale eddies to small-scale ones. \cite{2005MNRAS.356..778S} has calculated the IGM thermal history including these processes, and they have shown that IGM gas temperature is increased up to $\sim 10^4\,$K with nano-Gauss PMFs.

The 21-cm signal is a promising tool for probing baryon gas density and temperature at high redshifts. Considering the above effects of PMFs on the IGM gas physics, \cite{2008PhRvD..78h3005S, 2009ApJ...692..236S} have predicted the 21-cm signal with PMFs. \cite{2019MNRAS.488.2001M} has taken into account the decrease in energy of PMFs and put a constraint on PMF strength from the measurement of the 21-cm global signal with \textit{EDGES} \citep{Bowman_2018}. Future experiments of the 21-cm global signal may improve understanding of PMFs.

On the other hand, in a colder IGM enabled by dark matter-baryon interactions \citep{2018Natur.555...71B, munoz18}, the residual free electron fraction and Compton coupling are both suppressed \citep{2020PhRvD.102h3502D}. These effects significantly enhance the PMF-induced heating rates (via ambipolar diffusion and decaying turbulence) compared to the standard cosmological scenario \citep{2005MNRAS.356..778S, 2015MNRAS.451.2244C, 2019MNRAS.488.2001M, 2020MNRAS.498..918B}. Consequently, even relatively weak PMFs (with present-day strength $B_0 \lesssim 0.5 nG$) can transfer a substantial fraction of their magnetic energy into heating the IGM. This rapid energy transfer causes the PMF to decay faster than the conventional $(1 + z)^2$ scaling during the dark ages and cosmic dawn, indicating that PMFs are unlikely to be the primary cause of the later heating require to raise the temperature above CMB, though they shape early thermal history \citep{2020MNRAS.498..918B, 2019MNRAS.488.2001M}.

Moreover, when dark matter-baryon interactions and PMFs act together, a plateau-like feature appears in the redshift evolution of the IGM temperature because of the interplay between cooling from DM-baryon scattering and the PMF heating over a range of $z\sim50$-$150$ \citep{2020MNRAS.498..918B}. Importantly, the upper limits on allowed $B_0$ depend sensitively on the DM interaction strength $\sigma$ and DM particle mass $m_\chi$. While PMFs stronger than $\sim 0.1$\,nG are disallowed in standard models without DM-baryon scattering \citep{2019MNRAS.488.2001M}, values up to $\sim 0.4$\,nG become permissible if DM-baryon scattering is effective (for suitable combinations of $m_\chi$ and $\sigma$) \citep{2020MNRAS.498..918B}. However, PMFs $\gtrsim 1$\,nG remain unlikely, as they would require unrealistically strong cooling to offset the magnetic heating. Thus, the joint effects of PMFs and DM-baryon interactions significantly alter PMF evolution and the resulting IGM temperature history, with strong implications for interpreting the global 21-cm signal \citep{2019MNRAS.488.2001M, 2020MNRAS.498..918B, 2025JCAP...01..089B}.

Not only is the global signal of the 21-cm line from the cosmic dawn affected by PMFs, but the 21-cm power spectrum is also affected \citep{2005MNRAS.356..778S,2006MNRAS.372.1060T,2009ApJ...692..236S,2019JCAP...01..033K}. Since both the matter fluctuations and the temperature fluctuations affect the inhomogeneous 21-cm signal, their contributions are highly nonlinear effects. Therefore, predicting the PMF constraint from the 21-cm power spectrum needs careful discussion. The 21-cm power spectrum from the pre-reionization era with SKA might improve bounds on the PMFs $\lesssim 0.01 - 0.1$ nG \citep{2024PhRvD.109b3518C, 2025JCAP...01..089B}.

\subsection{Baryon -- Dark Matter Interactions}
A number of exotic physical models were stimulated by the possible EDGES detection of a strong 21-cm signal during cosmic dawn \citep{Bowman_2018}. While disputed at 95\% significance by the SARAS experiment \citep{SARAS3}, with further measurements expected to resolve this tension, the tentative EDGES signal has inspired theories that can be probed over a wide range of possible parameters, independently of whether EDGES turns out to be correct. 

One category of explanations for EDGES is the combination of Ly$\upalpha$ coupling from early galaxies with a mechanism that cooled the gas faster than just adiabatic cooling due to the cosmic expansion \citep{Barkana:2018}. The additional cooling mechanism that was suggested \citep{2014PhRvD..89b3519D,2014PhRvD..90h3522T,2015PhRvD..92h3528M,munoz17} involves a non-gravitational interaction between the ordinary matter and the dark matter particles (e.g., via Rutherford-like scattering); this drives down the gas temperature, leading to the strong observed absorption.

Any particle physics model that supplies such a new scattering interaction faces additional constraints \citep{Berlin, Barkana2018a}, including those on dark matter self-interaction (constrained by the known distribution of dark matter around galaxies) as well as baryon self-interaction, i.e., a fifth force (constrained by laboratory measurements). A model that satisfies these constraints is millicharged dark matter, in which a small fraction of the dark matter particles have a tiny electric charge, and Coulomb scattering is responsible for the energy transfer \citep{munoz18, munoz18b}. With any baryon -- dark matter interaction, the strong correlation between baryon temperature and the baryon -- dark matter relative streaming velocity \citep{TseliakHirat} tends to imprint large velocity acoustic oscillations on the 21-cm signal \citep{Barkana:2018}. However, this signature is erased in the millicharged dark matter model by drag at early times, throughout the viable parameter space of this model \citep{Kovetz}. It is possible to restore this signature in an interacting millicharged dark matter model \citep{Liu19,imDM}, which is more elaborate (adding a long-range interaction between the millicharged part and the rest of the dark matter) but also is viable over a much wider range of parameters.

These exotic models predict 21-cm signals that, over a significant range of model parameters, are substantially stronger than in any standard astrophysical model. Thus, the exotic models will be some of the earliest models to be constrained by observations, including those by the SKA. Figure~\ref{fig:RadioDM} shows an estimate of the maximum 21-cm power spectrum for the interacting millicharged dark matter model at wavenumber $k=0.13$~Mpc$^{-1}$ (right panel, solid blue curve). Interestingly, the maximum power spectrum from this model is comparable to the maximum power spectrum from another major category of models inspired by the EDGES signal, namely, strongly-emitting radio galaxies (Sec.~5.1). In both cases, the exotic model can yield a 21-cm power spectrum that is 3 orders of magnitude higher than the power
spectrum of standard astrophysical models, and 5 orders of magnitude above the expected SKA sensitivity.

\subsection{Dark matter annihilation/decay}\label{sec:DM-ann}
The fundamental nature of dark matter (DM) has troubled physicists ever since its existence was confirmed since the early 1970s \citep{Rubin_1970}. Past efforts on both the fronts, theory and observation have shown us that on cosmological scales, DM forms the skeleton of the Universe as seen today, i.e., the large-scale structure of our Universe \citep{Peacock_2001}. The currently accepted standard model assumes that DM is cold (CDM) and collisionless, and to a good approximation any non-gravitational interactions of DM are tiny. Yet there are several studies \citep[e.g.][]{Acevedo_2024} which propose DM-ordinary matter interactions to address observational inconsistencies such as the ``core-cusp'' problem \citep{Blok_2010, Oh_2011}. We refer the interested readers to \citet{Bergstrom_2009}, \citet{Bertone_2018}, and \citet{Mahmoudi_2021} for a further reading on DM -- listing the possible DM candidates, and detection strategies for different types of DM candidates. See \citet{Slatyer_2024} to read more about the interacting dark matter scenario from the perspective of cosmology. 

Various DM particle models have been proposed, including Weakly Interacting Massive Particles (WIMPs), axions, dark photons, and primordial black holes (PBHs). These candidates are being probed through multiple approaches such as direct detection, collider experiments, and astrophysical observations.

DM particles may annihilate into other Standard Model particles, for example electron-position pairs, muon pairs, tau lepton pairs, bottom quark pairs, and photon pairs.
These annihilation processes inject additional energy into the IGM in the Universe \citep{Slatyer:2009yq,Slatyer:2015jla,Slatyer:2015kla}. Similarly, DM particles may spontaneously decay into lighter Standard Model particles, thereby also releasing energy into the IGM \citep{Chen:2003gz}. These energy injections can heat and ionize the IGM \citep{Liu_2020, Liu:2023fgu, qin24, sun25b}, altering its thermal history and inevitably leaving observable cosmological imprints on cosmological 21-cm signal. Conversely, given a dataset it is possible to derive constraints on the fundamental properties of DM \citep[e.g.][]{Mittal_pbh}.

Current constraints from cosmological and astrophysical probes have placed stringent limits on such energy injection processes. Analysis of CMB polarization anisotropy by the Planck collaboration has set upper limits on the dark matter annihilation efficiency reaching $\sim 3 \times 10^{-28}~\mathrm{cm^3\,s^{-1}\,Gev^{-1}}$ for the $\chi\chi \rightarrow \mathrm{e}^{+}\mathrm{e}^{-}$ channel and a decay lifetime limit of $\sim 10^{24}\,\text{s}$ \citep{Plankc2018VI}. Complementary constraints come from Fermi-LAT observations of gamma rays from galaxies and galaxy clusters, yielding an limit on annihilation cross-section $\left<\sigma v\right> \lesssim 3 \times 10^{-26}~\text{cm}^3\,\text{s}^{-1}$ for $m_{\chi} = 100\,\text{GeV}$ and on decay lifetime $\tau_\chi \gtrsim 10^{26}~\text{s}$ \citep{Ackermann_2015PhRvL}. Other well-motivated candidates like axions and dark photons also face distinct bounds from CMB and other cosmological probes. For detailed discussions, we refer readers to specialized papers \citep{Bondarenko:2020moh,Das:2024ebv,Nguyen:2025tkl}. However, probes like the CMB primarily investigate physical processes in the late Universe (from post-reionization to the present day), whereas the 21-cm signal has the potential to open a new window into the DM effects in the earlier Universe stages, i.e., Cosmic Dawn and EoR.

\begin{figure*}
    \centering
\includegraphics[width=0.438\linewidth,,trim=0.0cm 0.3cm 0.0cm 0.2cm, clip]{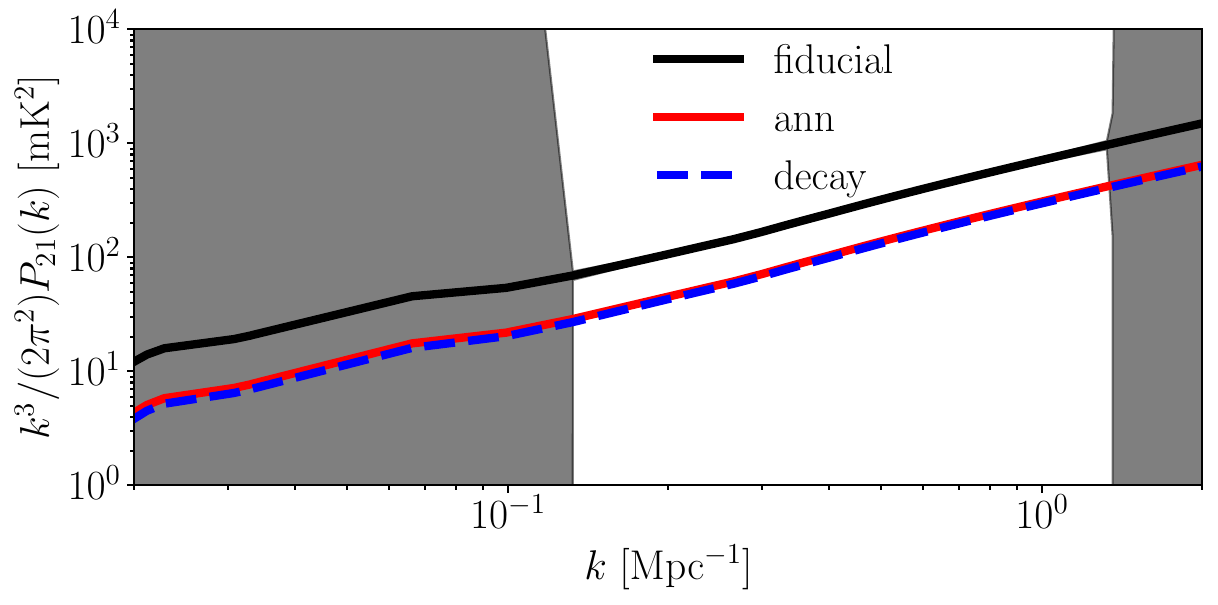}    \includegraphics[width=0.45\linewidth,,trim=0.0cm 0.3cm 0.0cm 0.2cm, clip]{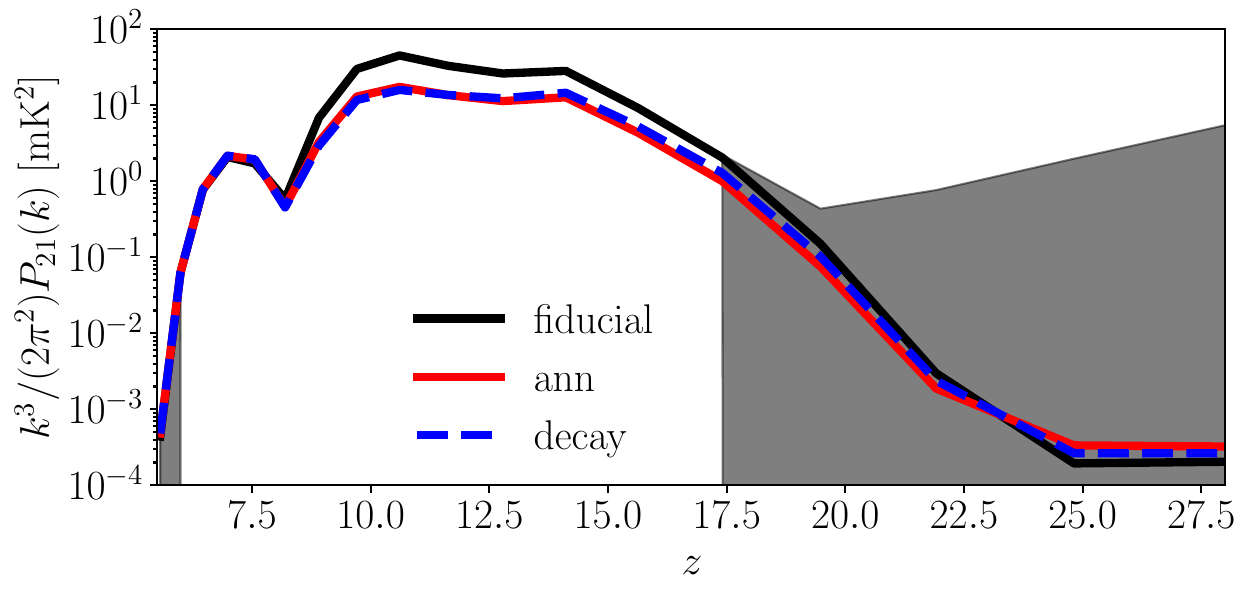}    
    \caption{
     {\it Left:} The 21 cm power spectra at $z\sim 10$ in the absence (black)/presence of DM annihilation (red) and decay (blue).
     {\it Right:} Similar to the left panel, however now the power spectra at $k=0.05$ Mpc$^{-1}$ as functions of redshift are plotted. 
     In all panels filled regions are SKA-AA* uncertainties for $t_{\rm obs}=1080$ hr and bandwidth 10 MHz,
      calculated by {\tt 21cmSense}, the left panel adopts moderate foreground removal model, while the right panel assumes optimistic case.
    }
    \label{fig:DM-21cmPS}
\end{figure*}

Fig.~\ref{fig:DM-21cmPS} shows the influence on 21-cm power spectra by annihilation or decay of DM particles with $m_\chi=100$ GeV. For annihilation, $\chi \chi \rightarrow \mathrm{e}^+\mathrm{e}^-$ channel and $\left<\sigma v\right>=3\times 10^{-26}{\rm\,cm^3\,s^{-1}}$ are assumed; for decay, $\chi \rightarrow \mathrm{e}^+\mathrm{e}^-$ channel and $\tau_\chi=10^{26}\,$s are adopted.
Compared with the observational uncertainties of SKA-AA*, the effects of DM are distinguishable at least between $k\sim 0.1 - 1$ Mpc$^{-1}$. Moreover, Cosmic Dawn is the optimal stage for detecting DM effects. As before this stage, the IGM has not yet well heated/ionized by DM, therefore the effects are small. After that, the heating and ionization of the IGM is dominated by first galaxies, the DM effects are negligible and are hard to see on the 21-cm signal.

Compared with other cosmological probes, the 21-cm signal is more sensitive to temperature variations in the IGM, enabling the detection of weaker energy injection processes beyond the reach of other probes. Moreover, the 21-cm power spectrum not only provides information about global evolution but also reveals spatial fluctuations, containing richer physical information that helps distinguish DM effects from astrophysical sources. Thus, future 21-cm observations by SKA are expected to provide unique tests of DM nature across different redshift and energy scales, thereby placing stringent constraints on the fundamental nature of DM particles.

\subsection{Non-cold dark matter}\label{sec:nonColdDM}


The first stars and galaxies in the Universe formed when baryons fell into gravitational potential wells created by bound dark matter particles. Consequently, the nature of dark matter significantly influences the properties of these luminous objects, including their typical masses and number densities. In this section, we consider a range of phenomenological dark matter models that affect the gravitational evolution of large-scale structure, such as warm dark matter \citep[WDM; e.g.,][]{sitwell2014imprint,giri2022imprints}, fuzzy dark matter \citep[FDM; e.g.,][]{fuzzy, nebrin2019fuzzy,jones2021fuzzy,giri2022imprints,flitter2022closing,sarkar2022exploring,2025PhRvD.112j3534L}, and the Effective Theory of Structure Formation \citep[ETHOS; e.g.,][]{Verwohlt:2024efh}. These models typically characterize dark matter particles by their rest mass energy, with more energetic particles suppressing the formation of small-scale structures. In particular, higher particle energies correspond to higher minimum halo masses capable of hosting luminous sources. In this work, we adopt the WDM scenario as a representative case for non-cold dark matter models.

\begin{figure*}
    \centering
    \includegraphics[width=0.8\linewidth,trim=0.0cm 0.0cm 0.0cm 0.0cm, clip]{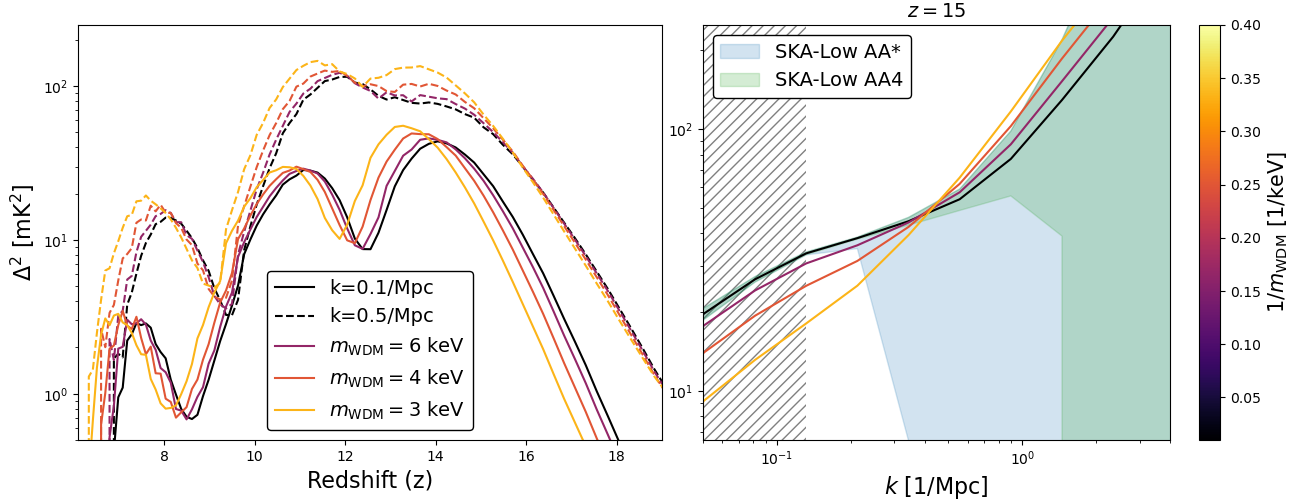}
    \caption{
    The power spectrum evolution over redshift (left panel) and wavenumber (right panel). The shaded regions correspond to the error due to cosmic variance and instrumental noise ($t_\mathrm{obs}=1000$h, $\Delta \ln k=0.5$) with AA* (blue) and AA4 (green) layout of SKA-Low.  
    }
    \label{fig:P21cm_WDM}
\end{figure*}

Previous studies have shown that these models are best distinguished at high redshifts during epochs such as cosmic dawn and reionization, where their impact on structure formation is most pronounced \citep[e.g.,][]{dayal2024warm}. Observations of the 21-cm signal by SKA-Low will provide a powerful probe of such models \citep[e.g.,][]{giri2022imprints}. Figure~\ref{fig:P21cm_WDM} illustrates the evolution of the 21-cm power spectrum for several WDM models, computed using the \textsc{HMreio} framework \citep{schneider2021halo,schneider2023cosmological}. The underlying astrophysical model is fixed to \textit{Model 1} in \citet{schneider2023cosmological}, with only the dark matter particle mass ($m_\mathrm{WDM}$) being varied. As $m_\mathrm{WDM}$ decreases, structure formation is progressively delayed, which in turn postpones the heating and reionization of the IGM, as seen in the left panel of Fig.~\ref{fig:P21cm_WDM}.
The right panel highlights the potential of SKA-Low to discriminate between these models via 21-cm power spectrum measurements at $z=15$. The shaded region around the CDM (cold dark matter) curve indicates the combined uncertainties from cosmic variance and instrumental noise. The hatched region marks scales $k \lesssim 0.15~\mathrm{Mpc}^{-1}$ that are heavily contaminated by foregrounds \citep{Pober_2014,Greig:2015qca}. Nevertheless, several foreground-free modes remain accessible for ruling out non-cold DM scenarios, even with a 1000-hour observation using the AA* layout. This capability improves substantially with the AA4 configuration, where a greater number of modes become available for discriminating among dark matter models. It is important to note that this sensitivity is redshift-dependent, as the signal-to-noise ratio varies across different epochs.

While our focus here has been on the 21-cm power spectrum, SKA-Low will be powerful enough to enable studies that go well beyond this two-point statistic \citep[e.g.,][]{giri2018optimal,giri2019neutral,giri2021measuring}. Several works have demonstrated that complementary observables—such as the 21-cm bispectrum \citep{saxena2020impact}, the 21-cm forest \citep{shimabukuro2014probing,shimabukuro2020constraining,shimabukuro202021,kawasaki2021probing,shao202321,sun2025deep}, and direct imaging \citep{neutsch2022inferring,sabiu2022probing}—can further constrain non-cold dark matter models. We refer interested readers to these studies for a broader perspective on how upcoming 21-cm observations may probe the fundamental nature of dark matter.

\subsection{Primordial features from inflation}
\label{subsec:inflation}
Inflation~\citep{Guth:1981,LINDE:1982,Albrecht:1982,STAROBINSKY198099,stato:1981,kazanas:1980} explains the origin of primordial density perturbations from quantum fluctuations of the early Universe. 
These fluctuations evolve to form large-scale structures and carry imprints of primordial physics. 
While current CMB and large-scale structure observations are broadly consistent with a nearly scale-invariant primordial power spectrum, deviations at certain scales may hint at new physics.
Several inflationary models predict scale-dependent features in the primordial power spectrum, or ``primordial features'' 
(see, e.g., \citealt{Akrami:2018odb, Chluba:2015bqa, Beutler:2019ojk} and references therein).
The neutral hydrogen distribution at high redshift carries imprints of primordial features through density fluctuation evolution. 
The 21~cm signal during cosmic dawn directly traces the matter distribution and is ideal for probing primordial features, though extraction is challenging due to parameter degeneracies. 
The tomographic nature of 21~cm observations helps disentangle such degeneracies~\citep{Naik:2022wej,Naik:2025mba}.

Previous studies have explored various types of primordial features with 21~cm intensity mapping and power spectrum, including oscillatory patterns from resonance or step-like features~\citep{Chen:2016zuu,Xu:2016kwz}, 
kink, step and warp shapes~\citep{Ballardini:2017qwq}, and dips from ultra-slow roll phases~\citep{Balaji:2022zur}. 
As a detailed case study, we focus on bump-like features arising from bursts of particle production during inflation  
\citep{Chung:1999ve,Barnaby:2009dd,Pearce:2017bdc}, which may occur naturally in models based on higher-dimensional gauge theories~\citep{Furuuchi:2015foh,Furuuchi:2020klq}. The primordial power spectrum is modeled by the height $A_{\rm I}$ and location $k_{\rm peak}$ of the bump~\citep{Pearce:2017bdc}.

\begin{figure}
    \centering
    \includegraphics[width=0.8\linewidth,trim=0.0cm 0.2cm 0.0cm 0.4cm, clip]{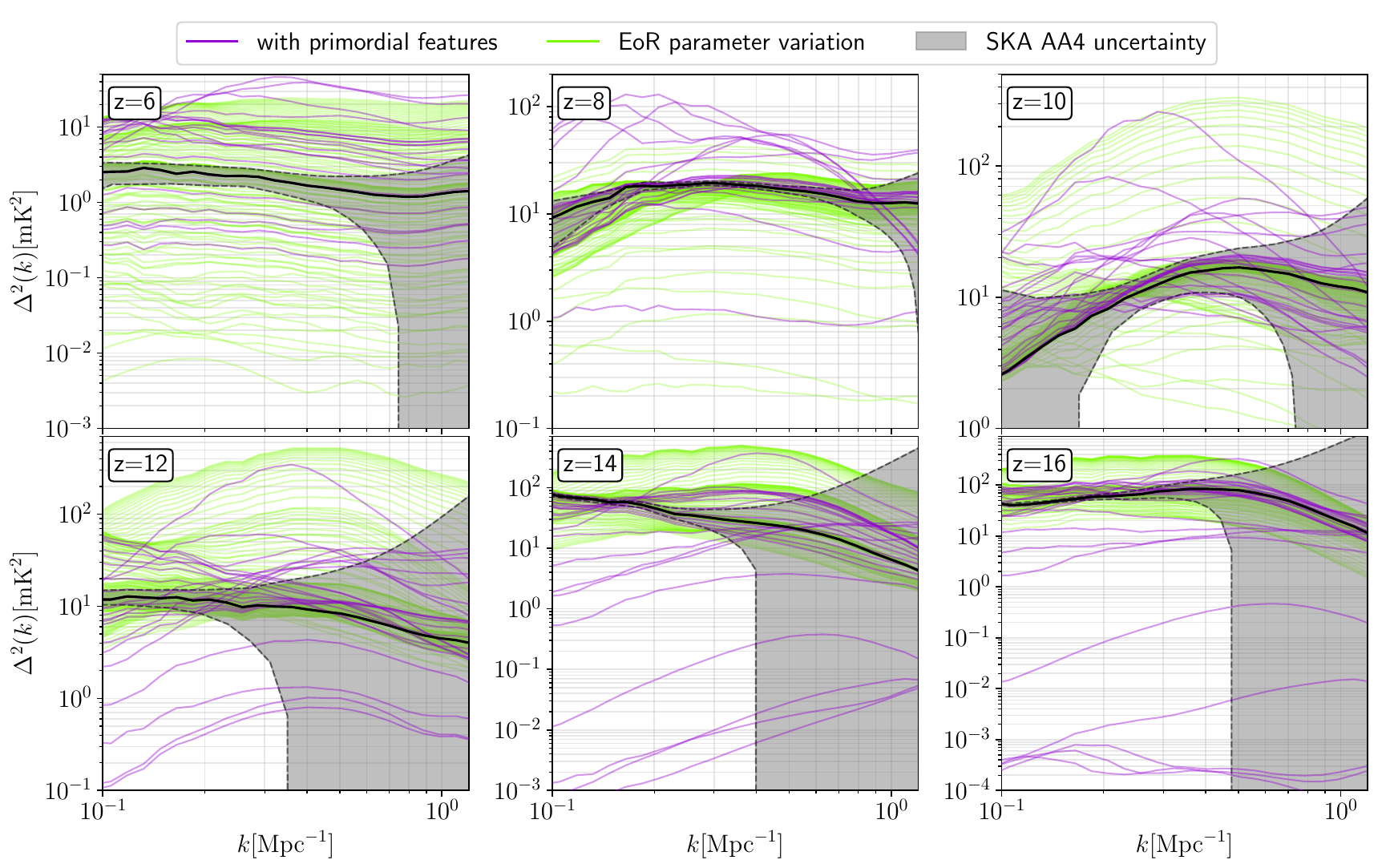}
    \caption{
The 21~cm power spectra at different redshifts simulated for the featureless power law model (solid black), models with bump-like features (violet) and various EoR models (green). The shaded region is the 95\% uncertainty estimated for SKA-AA4 using \texttt{21cmSense}.    
}
    \label{fig:prim_feature}
\end{figure}

Bump-like features enhance correlations in density fluctuations, significantly altering 21~cm brightness temperature fluctuations and power spectra (figure~\ref{fig:prim_feature}). 
These features also affect the reionization history and global 21~cm profile \citep{Naik:2025mba,Yoshiura:2018zts,Yoshiura:2019zxq}, primarily through modifications to the halo mass function. 
As $k_{\rm peak}$ increases, reionization shifts from late- to early-completing scenarios, affecting all observables~\citep{Naik:2025mba}.
Other types of primordial features are also expected to affect density fluctuations and the halo mass function. Detailed SKA forecasts for these feature types remain to be explored. 

In \cite{Naik:2022wej}, it was demonstrated that multi-redshift 21-cm power spectra can probe bump-like features within $0.1 \le k\,[{\rm Mpc}^{-1}] \le 1.0$ with SKA-Low earlier layout. 
Figure~\ref{fig:prim_feature} shows 21~cm power spectra for various bump model parameters (violet) and EoR models (green). The shaded regions indicate 95\% uncertainty at different redshifts for the SKA-AA4 layout. 
Simulations were performed using \texttt{21cmFAST}~\citep{Mesinger_11,Murray:2020trn} and 
the noise estimates were computed with \texttt{21cmSense} (\citealt{murray202421cmsense,Pober2013aj,Pober:2014apj}), assuming 6 hours per day of deep field observations for 180 days with optimistic foreground removal.  
The unique spectral shapes of the models with primordial features show strong $k$ and $z$ dependence. 
At certain scales and redshifts, primordial feature effects are quite distinct from EoR parameter variations. 
This demonstrates promising prospects for detecting primordial features with SKA-Low AA4 through combining multi-redshift 21~cm power spectra. 

\subsection{Other exotic signatures}
\label{sec:othersignatures}
Topological defects such as superconducting cosmic strings are predicted by several different extensions to the standard model. If these defects exist, radio emission from their oscillations, cusps and kinks has the potential to modify the 21-cm signal. This is both through direct enhancement of the radio background, and potentially also via soft photon heating \citep{acharya2023sph1,cyr2024sph2}; a poorly-studied yet likely non-negligible effect in models that include a large excess radio background. Current upper limits on the 21-cm power spectrum provide no observational constraints on the existence or properties of cosmic strings \citep{gessey2024constraints}, but it has been proposed that SKA 21-cm signal images or three-point statistics could reveal overdensity structures associated with cosmic strings \citep{hernandez2014cosmicstringwakes1,maibach2021cosmicstringwakes2}. 
\\
\\
The inverse Gertsenshtein effect (i.e., conversion of gravitational waves to photons in the presence of a magnetic field) has been posed as a mechanism for detecting high-frequency primordial gravitational waves with radio telescopes (e.g. \citealp{domcke2021potential}) via the resulting enhancement of the radio background. Forecasts suggest that the SKA may improve current MHz-band constraints on the primordial gravitational wave density parameter $\Omega_\text{GW}$ by 7 -- 10 orders of magnitude \citep{he2024inverse}. In principle, all processes that impact halo formation may also leave an imprint in the 21-cm power spectrum, since halo statistics affect galaxy formation, which enables the feedback processes described in Section \ref{sec:21cmsignal}. Some unmentioned examples of such processes include early/dynamical dark energy \citep{yin2023_dyn_lambda_21cm,adi2025_early_lambda_21cm}, free-streaming of massless \citep{montefalcone25} and massive neutrinos \citep{pritchard2008_massive_neutrinos,zhang2020_massive_neutrinos, libanore25} and isocurvature perturbations \citep{hotinli21, takeuchi2014_isocurvature_21cm,qin2025_isocurvature_21cm}.

\section{Conclusions}
In this chapter we have discussed the prospects of the upcoming low-frequency Square Kilometre Array telescope (SKA-Low) to probe standard and exotic signatures from the Cosmic Dawn and the Epoch of Reionization (EoR). The SKA-Low interferometer will be sensitive to frequencies between 50-350 MHz, allowing for a high-precision measurement of the redshifted 21-cm signal from the redshift range of $z\sim 6-30$.

The chapter is structured into two main sections discussing standard and exotic signatures from the Cosmic Dawn and the EoR. Among the standard astrophysical signatures, SKA-Low will enable detailed investigations of the first galaxies and their feedback on the IGM. Variations in star-formation efficiency, escape fraction, and duty cycle will modulate the amplitude and scale-dependence of the 21-cm power spectrum, providing constraints on early galaxy formation models beyond the reach of current optical surveys. The imprint of Population III stars, whose intense UV and X-ray radiation initiate the Ly$\upalpha$ coupling and early heating, will be reflected in both the timing and morphology of the 21-cm absorption trough. X-ray binaries, expected to dominate the high-energy emission during cosmic dawn, will drive the global transition from absorption to emission, while cosmic rays from early supernovae will produce a spatially extended heating signature detectable through a suppression of small-scale 21-cm power. The presence of supermassive black holes will generate large ionized bubbles and local heating patterns, imprinting scale-dependent anisotropies that SKA-Low can distinguish in its tomographic power-spectrum analysis. Finally, velocity acoustic oscillations (VAOs), arising from baryon-dark-matter streaming velocities, will both modulate star formation in mini-haloes and appear as periodic modulations in the 21-cm power spectrum providing a novel standard ruler for cosmology at high redshifts.

Beyond these standard astrophysical processes, SKA-Low will be sensitive to a wide range of exotic signatures that could reveal new physics beyond the $\Lambda$CDM paradigm. These include modifications of the 21-cm signal due to strong radio emitting galaxies or due to the energy injection induced by dark-matter interactions, annihilation or decay, primordial black holes, or primordial magnetic fields. The small-scale clustering could be altered by warm dark matter components, additional hot species, or explicit features imprinted by non-standard inflation. The detection or exclusion of such effects would provide unique constraints on the nature of dark matter, or reveal new and exotic sources.

In summary, the SKA-Low will transform our understanding of the high-redshift Universe and the complex astrophysical processes that shaped it, including the emergence of the very first luminous sources. At the same time, it will serve as a powerful instrument to search for new physics beyond the standard cosmological model, particularly in the dark matter sector. Together, these capabilities will make SKA-Low a cornerstone of early-Universe astrophysics and cosmology.

\section{Contributions}
A. Schneider coordinated the work on this science chapter and wrote abstract, introduction, and conclusion.
S. Mittal led Sec.~\ref{sec:Lyacoupling} and \ref{sec:DM-ann}, contributed to Sec.~\ref{sec:heating}, \ref{sec:XRBs}, \ref{sec:radiogal}, and homogenised the bibliography.
A. Fialkov led Sec. \ref{sec:heating} and \ref{sec:firststars}, and contributed to sections \ref{sec:XRBs} and \ref{sec:radiogal}.
I. T. Iliev led Sec.~\ref{sec:ion_process}, contributed to Sec. 2.1, 2.2, 4.2 and 4.6.
J.Y.H. Chan led Sec.~2.4. 
S. K. Giri led Sec.~\ref{sec:SKALow} and \ref{sec:nonColdDM}.
R. Barkana led Sec.\ 4.3, 4.7, and 5.3.
A. Bera led Sec. \ref{sec:CRs} and contributed to Sec. \ref{sec:PMFs}.
B. Yue led Sec. \ref{sec:SMBHs}  and  Sec. \ref{sec:early-galaxies}, and contributed to Sec. \ref{sec:DM-ann}.
H. A. G. Cruz and J. Mu\~noz led Sec.~\ref{sec:vao} and produced Fig.~\ref{fig:VAOs}.
B. Li co-led Sec.~\ref{sec:other} and contributed to Sec.~\ref{sec:nonColdDM}.
S. Sikder led Sec.\ 5.1 and made Fig.\ 5.
T. Minoda led Sec. \ref{sec:PMFs}. 
S. S. Naik led sec.~\ref{subsec:inflation} and produced Fig.~\ref{fig:prim_feature}.
O. Basquette led Sec. \ref{sec:othersignatures}, contributed to Sec. \ref{sec:firststars} and produced Fig. \ref{fig:ps_IMF_constraints}.
S. Dasgupta produced Fig.\ 1.
J. Raste contributed to Secs. \ref{sec:21cmsignal}, \ref{sec:Lyacoupling}, and \ref{sec:heating}.
Q. Han and K.Wan contributed to Sec.~2.4.
K. K. Datta contributed to Sec. \ref{sec:CRs}.
M. Zhang contributed to Sec. \ref{sec:SMBHs}.
Y. Xu contributed to Sec. \ref{sec:SMBHs} and Sec. \ref{sec:DM-ann}.
M.-L. Zhao contributed to Sec. \ref{sec:DM-ann}.
P. Chingangbam contributed to Sec.~\ref{subsec:inflation}.

\section{Acknowledgments}

RB and SS acknowledge the support of the Israel Science Foundation (grant no.\ 1078/24).
YX acknowledges the support from the National SKA Program of China No. 2020SKA0110401, and the National Key R\&D Program of China No. 2022YFF0504300.
BY acknowledges the support from the National SKA Program of China No. 2020SKA0110402, and the NSFC International (Regional) Cooperation and Exchange Project No. 12361141814. 
SM is supported by the ERC (UKRI guaranteed) research grant EP/Y02916X/1.

\bibliographystyle{abbrvnat-maxbibnames4}
\bibliography{biblio} 

\end{document}